%% file: main.tex
\documentclass[journal=jacsat, manuscript=article, layout=twocolumn]{achemso}
\usepackage{graphicx}
\usepackage{amsmath}
\usepackage{bm}
\usepackage{float}
\usepackage{url}
\usepackage{placeins}

\makeatletter
\renewcommand\normalsize{%
    \@setfontsize\normalsize{10pt}{12pt}%
}
\makeatother

\title{Inverse Design Method with Enhanced Sampling for Complex Open Crystals: Application to Novel Zeolite Self-Assembly in a Coarse-Grained Model}

\author{Chaohong Wang}
\email{c.wang2@uu.nl}
\affiliation{Soft Condensed Matter \& Biophysics, Debye Institute for Nanomaterials Science, Utrecht University, Princetonplein 1, 3584 CC Utrecht, The Netherlands}

\author{Alberto P\'erez de Alba Ort\'iz}
\email{a.perezdealbaortiz@uva.nl}
\affiliation{Soft Condensed Matter \& Biophysics, Debye Institute for Nanomaterials Science, Utrecht University, Princetonplein 1, 3584 CC Utrecht, The Netherlands}
\affiliation{Computational Soft Matter, van 't Hoff Institute for Molecular Sciences and Informatics Institute, University of Amsterdam, Science Park 904, 1098 XH Amsterdam, The Netherlands}

\author{Marjolein Dijkstra}
\email{m.dijkstra@uu.nl}
\affiliation{Soft Condensed Matter \& Biophysics, Debye Institute for Nanomaterials Science, Utrecht University, Princetonplein 1, 3584 CC Utrecht, The Netherlands}

\begin{document}

\begin{abstract}
    Optimizing the synthesis of zeolites and exploring novel frameworks offer pivotal opportunities and challenges in materials design. While inverse design proves highly effective for simpler crystals, its application to intricate structures like zeolites poses severe challenges. Here, we introduce an innovative inverse design workflow tailored to efficiently reproduce target zeolite frameworks in a binary coarse-grained model using enhanced sampling molecular dynamics simulations. This workflow integrates an evolutionary parameter optimization strategy with a variant of the seeding approach. Using this method, we successfully reproduce Z1 and SGT zeolites, and Type-I clathrates, find new optimal parameters for known phases, such as the SOD and CFI, and even discover novel frameworks, such as Z5. This is done within a simple coarse-grained model for a tetrahedra-forming component and a structure-directing agent. Our methodology not only enables the screening of synthesis protocols but also facilitates the discovery of hypothetical zeolites.
\end{abstract}

\input{introduction}

\input{method}

\section{Results}

\input{results_z1}

\input{results_sgt}

\input{results_sod}

\input{results_cfi}

\input{results_afi}

\input{conclusion}

\begin{acknowledgement}
    CW, APAO and MD gratefully acknowledge SURF (www.surf.nl) for the computational resources, the national supercomputer Snellius, and is also grateful for the financial support from the Chinese Scholarship Council (CSC). APAO and MD acknowledge funding from the European Research Council (ERC) under the European Union’s Horizon 2020 research and innovation program (Grant agreement No. ERC-2019-ADG 884902, SoftML).
\end{acknowledgement}

\bibliography{zeolites}
\end{document}

%% file: introduction.tex
\section{Introduction}

Zeolites are microporous aluminosilicate crystals characterized by open three-dimensional framework structures composed of corner-sharing ${\rm TO_4}$ tetrahedra, where T typically represents silicon (Si) or aluminum (Al) in varying ratios. Zeolites are extensively used in numerous industrial and technological applications, including as catalysts~\cite{weitkamp2000zeolites, corma1995inorganic, corma1997microporous}, molecular sieves~\cite{corma1997microporous,sherman1999synthetic}, adsorbents for environmental protection~\cite{wang2010natural}, and even as biomedical materials~\cite{serati2020biomedical}. This variety of applications is enabled by the vast diversity of specific pore shapes and sizes with which zeolites can be synthesized. To produce such zeolites, hydrothermal synthesis (HS) is the most common route~\cite{cundy2005hydrothermal}. In this method, a precursor such as tetraethylorthosilicate (TEOS) is mixed with water and a so-called structure directing agent (SDA), typically a tetraalkylammonium cation or another organic cation, under specific temperature and pressure conditions. The SDAs interact with the framework precursors, playing a crucial role in guiding the polymerization, self-aggregation, crystallization, and growth of the zeolite~\cite{lobo1995structure}. By adjusting the composition of T atoms and SDAs, as well as the thermodynamic conditions, one can control the metastability of different polymorphs and promote the formation of a specific zeolite framework. Hundreds of synthetic and natural zeolite frameworks have been cataloged by the Structure Commission of the International Zeolite Association (IZA-SC)~\cite{DatabaseofZeoliteStructures}. Still, more than four million hypothetical zeolites remain unrealized in experiments, according to databases built by varying unit cell parameters, density, T atom positions and compositions using Monte Carlo search methods~\cite{earl2006toward, deem2009computational, pophale2011database}.

The potential applications of these hypothetical zeolites drive the search for novel strategies to fine-tune hydrothermal synthesis. Producing zeolites with desired properties, including hypothetical ones, requires identifying the appropriate T elements and their ratios, as well as optimizing the thermodynamic conditions, and selecting the optimal SDAs in the most effective compositions. These agents are crucial for guiding the framework components to form specific topologies. In this endeavor, computational approaches have become increasingly crucial for understanding and controlling zeolite self-assembly. Molecular dynamics (MD) simulations have provided insights into the zeolite self-assembly process, while artificial intelligence (AI) and optimization algorithms have advanced the design of stable zeolite frameworks. Here, we combine accelerated MD simulations with evolutionary computing techniques to facilitate the self-assembly of target frameworks within a coarse-grained zeolite model.

Efficiently screening for optimal conditions for the self-assembly of a target zeolite framework is prohibitively expensive when relying on systematic experimentation or forward modeling. The exploration of such combinatorial design parameter spaces has motivated the surge of inverse design methods.
Inverse design is a bottom-up approach that involves adjusting the properties of building blocks, such as particle shape and particle interactions, as well as the thermodynamic conditions, to achieve the desired properties of the self-assembled material. Many inverse design frameworks have been proposed in recent decades~\cite{torquato2009inverse}. For instance, Torquato and coworkers used inverse statistical mechanical methods to develop isotropic potentials that facilitate the formation of various colloidal lattices~\cite{rechtsman2005optimized, torquato2009inverse, marcotte2013communication}. Other researchers have also designed potentials and conditions to identify quasicrystals through inverse design techniques~\cite{coli2022inverse, bedolla2024inverse}. However, porous materials such as zeolites and metal-organic frameworks (MOFs) pose unique challenges for inverse design due to their relatively complex topologies compared to simpler crystalline materials. Some studies have identified novel thermodynamically stable porous materials by leveraging existing data sources, such as the IZA-SC, and employing artificial intelligence, without considering the self-assembly mechanisms. For instance, porous materials have been designed using generative adversarial neural networks~\cite{kim2020inverse, zheng2023deep}, variational autoencoders~\cite{yao2021inverse}, diffusion models~\cite{park2024inverse}, and other machine learning methods~\cite{gandhi2022machine}. While previous work has employed optimization strategies to design the self-assembly of open crystals~\cite{pineros2017designing}, to the best of our knowledge, a generic, robust inverse design framework capable of optimizing the self-assembly of complex zeolitic frameworks does not currently exist.

An iterative inverse-design framework for zeolite self-assembly requires computational efficiency that is difficult to achieve with atomistic models. Recently, Molinero {\em et al.} proposed a novel and concise coarse-grained model for zeolites consisting of tetrahedral network-former T particles and structure-directing agent S particles~\cite{kumar_two-step_2018}. The T particles are parameterized based on the mW water model~\cite{molinero2009water}, which in turn is a special parametrization of the Stillinger-Weber (SW) model~\cite{stillinger1985computer}. The T-S model considers two-body interactions between T-T, T-S and S-S pairs, as well as an additional three-body interaction between T particles to capture tetrahedrality. This coarse-grained T-S model has been reported to reproduce several known structures, including the SGT zeolite framework and the sII clathrate, and a zeolite analog of the FIR-30 metal-organic framework dubbed Z1~\cite{kumar_two-step_2018, dhabal_coarse-grained_2021, wang_homochiral_2015}, along with various other phases~\cite{kumar_could_2018}. Due to the complexity of phases introduced by many-body interactions, binary components, and a broad range of adjustable interaction parameters, this model has the potential to reproduce a wide diversity of zeolite frameworks~\cite{dhabal_what_2022}. The T-S model is also highly tunable, with free parameters such as particle size, interaction strength and range, tetrahedrality, making it well-suited for the inverse design of the self-assembly of desired zeolites.

In this work, we devise an innovative inverse design workflow capable of identifying optimal parameter values in a multidimensional parameter space for the self-assembly of desired zeolite frameworks. Our workflow utilizes the coarse-grained T-S model to represent interactions between T and S particles and to propagate their dynamics. To determine promising parameter values in the T-S zeolite model, we employ an evolutionary approach --- the covariance matrix adaptation evolution strategy (CMA-ES)~\cite{hansen2001completely}--- which is a robust, population-based, gradient-free optimizer. The parameters we optimize correspond to the effective size and the attraction strength between T-T, T-S and S-S pairs of particles, the strength of the tetrahedrality in T particles, and the temperature. To ensure that the identified potential parameter values can lead to self-assembly into the target framework, the fitness of the optimizer is determined from a nucleation perspective. To accelerate the nucleation and growth of our framework even further, we employ enhanced sampling techniques, i.e., seed-pinning~\cite{bai_calculation_2006, espinosa2016seeding}, to favor nucleation events that are typically rare in standard MD simulations. To this end, we introduce a seed, i.e., a small crystallite of the desired zeolite into a fluid mixture and monitor its growth to evaluate the fitness of the parameters. To determine this tendency to grow, we measure fluctuations in the environment similarity order parameter~\cite{piaggi_calculation_2019} with respect to the target framework as cataloged in the IZA-SC database. Additionally, we implemented a novel algorithmic approach to reduce the computational cost of sampling and measuring nucleus size fluctuations, thereby increasing the efficiency of the inverse design.

This paper is organized as follows. In Section II, we describe each of the methodology's elements, including the T-S zeolite model, the inverse design workflow, the CMA-ES optimizer, the environment similarity order parameter, and the seed-pinning method. In Section III, we present our results for the Z1 and SGT zeolite frameworks, as well as newly discovered parameters for SOD, sI clathrate, CFI, and the new framework Z5. Finally, we provide our conclusions in Section IV.

%% file: method.tex
\section{Methods}

In this section, we describe the key elements of our inverse design workflow. First, we present the coarse-grained T-S model used to determine particle interactions and compute forces for running molecular dynamics (MD) simulations. This model is parameterized by the range and strength of interactions, as well as the strength of tetrahedrality and temperature. Second, we explain how these model parameters are iteratively optimized using the covariance matrix adaptation evolution strategy (CMA-ES). Since CMA-ES requires a fitness function to guide the optimization, we show how we measure proximity to a target zeolite structure using the environment similarity order parameter~\cite{piaggi_calculation_2019}. To further speed-up the inverse-design pipeline, we introduce a seed-pinning technique that accelerates the sampling of nucleation events, allowing us to rapidly assess fluctuations in the environment similarity order parameter-based fitness. Finally, we provide specific details on the simulation parameters and protocol.

\subsection{Zeolite model}
The coarse-grained zeolite model proposed by Molinero {\em et al.} is a binary mixture consisting of T and S particle types. The T particles, which exhibit tetrahedral interactions, represent the tetrahedrally coordinated atoms, such as silicon (Si), aluminum (Al), or other heteroatoms. The S particles, on the other hand, represent the structure-directing agent (SDA), typically an organic cation. The interactions between these particles are described by the Stillinger–Weber (SW) potential~\cite{stillinger1985computer}, and the potential energy $U$ of the system reads
\begin{equation}
    U = \sum_{i < j} \phi^{(2)}_{\alpha\beta}\left(r_{i j}\right) + \hspace{-3mm}\sum_{i,j \neq i, k>j,k \neq i} \phi^{(3)}_{\mathrm{TT}}\left(r_{i j}, r_{i k}, \theta_{i j k}\right),
\end{equation}
where $r_{i j}=|{\bf r}_{ij}|$ is the distance between particle $i$ and $j$ with ${\bf r}_{ij}={\bf r}_i-{\bf r}_j$, $\theta_{i j k}$ is the angle between ${\bf r}_{i j}$ and ${\bf r}_{i k}$, and $\phi^{(2)}_{\alpha\beta}$ represents the two-body interaction applied to both T and S species 
\begin{eqnarray}
    \phi^{(2)}_{\alpha\beta}\left(r_{i j}\right)&=& \nonumber \\
    & & \hspace{-15mm} A \varepsilon_{\alpha\beta}\left[B\left(\frac{\sigma_{\alpha\beta}}{r_{i j}}\right)^p-\left(\frac{\sigma_{\alpha\beta}}{r_{i j}}\right)^q\right] \varphi_{\alpha\beta}(r_{i j}),
    \label{eq_two-body}
\end{eqnarray}
where $\alpha\beta$ denotes the particle pair types TT, TS, or SS, and  where 
\begin{equation}
    \varphi_{\alpha\beta}\left(r_{i j}\right)=\exp \left(\frac{\sigma_{\alpha\beta}}{r_{i j}-a \sigma_{\alpha\beta}}\right).
\end{equation}
The three-body interaction $\phi^{(3)}_{\mathrm{TT}}$ applied to solely the T particle species is denoted by 
\begin{eqnarray}
    \phi^{(3)}_{\mathrm{TT}}\left(r_{i j}, r_{i k}, \theta_{i j k}\right)&=& \nonumber \\
    & & \hspace{-25mm}\lambda \varepsilon_{\mathrm{TT}}\left(\cos \theta_{i j k}-\cos \theta_0\right)^2 \psi(r_{i j}) \psi(r_{i k}),
\label{eq_tetrahedral}
\end{eqnarray}
with
\begin{equation}
    \psi\left(r_{i j}\right)=\exp \left(\frac{\gamma \sigma_{\mathrm{TT}}}{r_{i j}-a \sigma_{\mathrm{TT}}}\right).
\end{equation}
The three-body interaction term favors the formation of a tetrahedral structure when $\theta_0=109.47^\circ$. 
The exponential terms $\varphi$ and $\psi$ ensure that both the potential and its derivatives smoothly approach zero at a cut-off distance of $r=a\sigma$. The tetrahedrality parameter $\lambda$ determines the strength of the tetrahedral interaction. In our work, the values of $A$, $B$, $p$, $q$, $\gamma$, and $a$ are consistent with those in the original SW potential~\cite{stillinger1985computer}, shown in Table~\ref{tab:sw}.

To achieve an average T-T bond length that matches closer to the Si-Si distance in zeolites and also aligns with Molinero's recent works~\cite{bertolazzo_unstable_2022, chu2024interzeolite}, we use the value of $\sigma_{\mathrm{TT}} = 2.7275$ \AA. The values of $\varepsilon_{\mathrm{TT}}$ and $\lambda$ are related to the elements and ratios of the framework components. Similarly, $\varepsilon_{\mathrm{SS}}$ and $\sigma_{\mathrm{SS}}$ depend on the SDA. The values of $\varepsilon_{\mathrm{TS}}$ and $\sigma_{\mathrm{TS}}$ reflect the interaction strength and range between the framework components and the SDA. Therefore, the design parameters include $\varepsilon_{\mathrm{TT}}$, $\varepsilon_{\mathrm{TS}}$, $\varepsilon_{\mathrm{SS}}$, $\sigma_{\mathrm{TS}}$, $\sigma_{\mathrm{SS}}$, $\lambda$, and temperature $T$.

\begin{table}[]
    \setlength{\tabcolsep}{3pt}
    \centering
    \begin{tabular}{cccc}
        \hline
        Parameter & Value & Parameter & Value \\\hline
        $A$ & 7.049556277 & $B$ & 0.6022245584 \\
        $p$ & 4 & $q$ & 0 \\
        $\gamma$ & 1.2 & $a$ & 1.8 \\\hline
    \end{tabular}
    \caption{Interaction parameters of the SW potential~\cite{stillinger1985computer} described by $\phi^{(2)}_{\alpha\beta}$ in Eq. (\ref{eq_two-body}) and $\phi^{(3)}_{\mathrm{TT}}$ in Eq. (\ref{eq_tetrahedral}).}
    \label{tab:sw}
\end{table}

\subsection{Inverse design workflow}
\begin{figure}[H]
    \centering
    \includegraphics[width=1.0\linewidth]{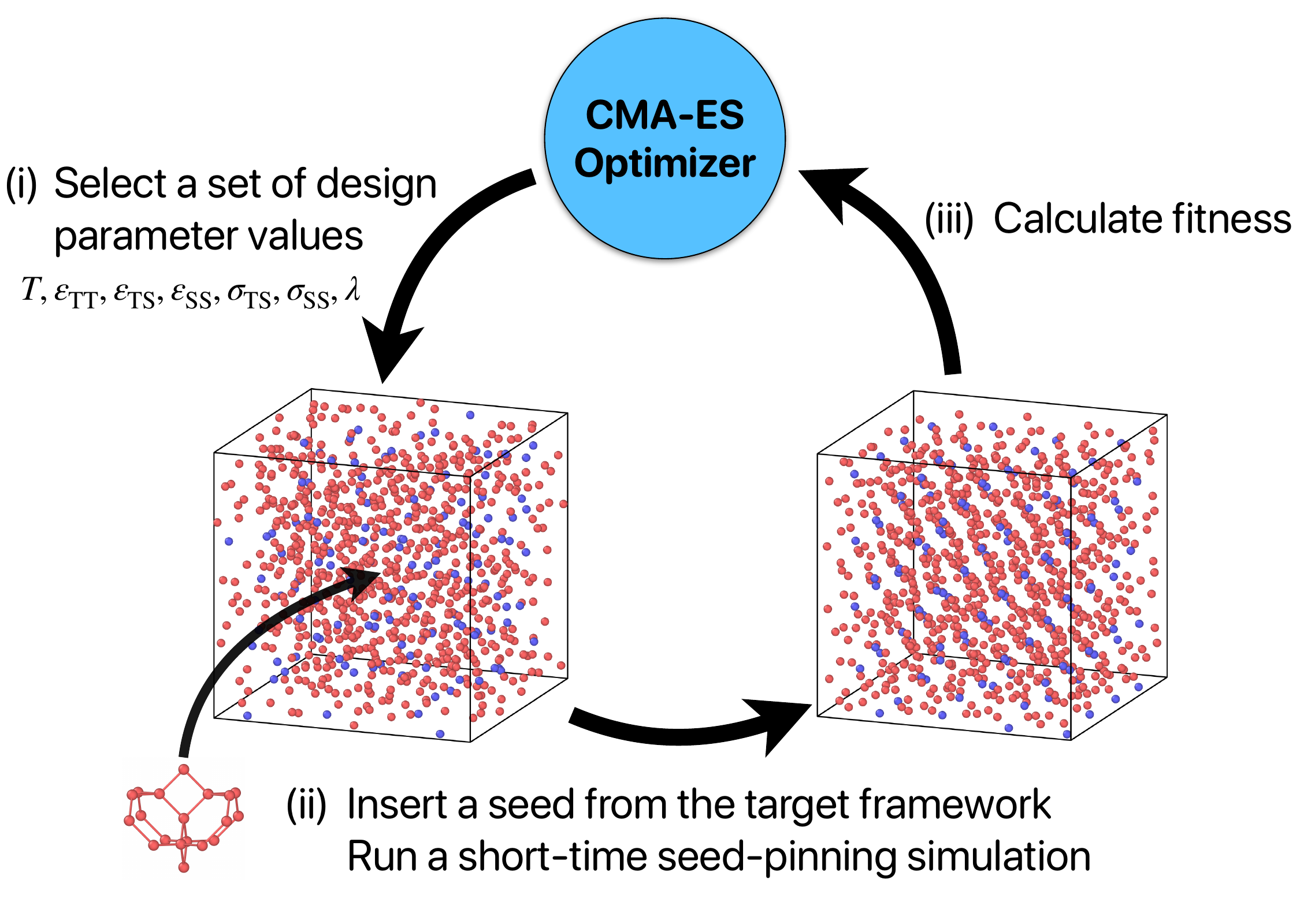}
    \caption{Schematic of the inverse design method. The workflow consists of the following steps: (i) selecting a set of design parameter values using the CMA-ES optimizer; (ii) inserting a seed of the target zeolite into a fluid phase and initiating a short-time molecular dynamics (MD) simulation, using the seed-pinning method; (iii) determining the fitness value by monitoring the environment similarity with respect to the target zeolite during the MD simulations. This fitness value is then fed back into the CMA-ES optimizer, which refines the design parameter values for the next iteration.
     }
    \label{fig:workflow}
\end{figure}

To optimize the design parameter values of the coarse-grained T-S model for the self-assembly of a target zeolite, we implement an iterative process described in this Section. Our inverse design workflow involves three key steps: (i) selecting a set of design parameter values from the CMA-ES optimizer (see Section~\ref{subsec:cmaes}); (ii) initiating an MD simulation in the disordered fluid phase with these parameters, using the seed-pinning method to accelerate the nucleation and growth of the target phase (see Section~\ref{subsec:seed-pinning}); and (iii) calculating the fitness value from the resulting MD trajectory---specifically, the environment similarity order parameter relative to the target zeolite---and feeding this information back to the optimizer to refine the design parameter values for the next iteration. In this study, we choose as design parameters, temperature $T$, interaction strengths between different species $\varepsilon_{\mathrm{TT}}$, $\varepsilon_{\mathrm{TS}}$, $\varepsilon_{\mathrm{SS}}$, interaction ranges between different species $\sigma_{\mathrm{TS}}$, $\sigma_{\mathrm{SS}}$, and the strength of tetrahedrality $\lambda$. The following subsections provide an in-depth explanation of these three steps, as schematically illustrated in Fig.~\ref{fig:workflow}.

\subsection{Covariance matrix adaptation evolutionary strategy}\label{subsec:cmaes}
In this work, we use the covariance matrix adaptation evolution strategy (CMA-ES) to optimize the fitness value, which measures how closely the system resembles the target phase. CMA-ES is a well-established, population-based, gradient-free stochastic optimization algorithm for real-valued, non-convex, and nonlinear functions~\cite{hansen2001completely}. The method operates without requiring gradient information and has been applied in inverse design studies~\cite{kumar_assembly_2019,coli2022inverse,bedolla2024inverse}.

CMA-ES is an iterative algorithm that samples from a multivariate Gaussian distribution and adapts the mean vector and covariance matrix at each iteration to reach the optimal solution. In each iteration, often referred to as a generation, the algorithm draws $n$ samples from a $d$-dimensional multivariate Gaussian distribution, where $d$ represents the number of design parameters. The fitness function is then evaluated for these samples, and the outcomes are ranked in descending order. The top $k$ samples are selected as the best candidates and form the set $\boldsymbol{X}$. Based on these best candidates, the algorithm updates the mean vector $\boldsymbol{m} \in \mathbf{R}^{d}$ and the covariance matrix $\boldsymbol{C} \in \mathbf{R}^{d \times d}$ to adjust the Gaussian distribution toward the optimal solution. This iterative process continues until a convergence criterion is met. Below is a brief description of each step in an iteration.

\begin{enumerate}
    \item A population of $n=4+\left\lfloor 3 \ln d\right\rfloor$ random samples~\cite{loshchilov2014maximum} is drawn from a normal distribution, where $d$ represents the dimension of the parameter space, i.e., the number of design parameters. Each sample or individual is generated as $\boldsymbol{x}_{i}^{(g+1)} \sim \mathcal{N}(\boldsymbol{m}^{(g)}, (\sigma^{(g)})^2 \boldsymbol{C}^{(g)})$, where $\boldsymbol{m}^{(g)}$ is the mean vector, $\sigma^{(g)}$ is the step size, and $\boldsymbol{C}^{(g)}$ is the covariance matrix at generation $g$.
    \item Each sample $\boldsymbol{x}_{i}^{(g+1)}$ is evaluated using the fitness function $f(\boldsymbol{x}_{i}^{(g+1)})$, and the samples are ranked according to their fitness values.
    \item The mean vector $\boldsymbol{m}^{(g+1)}$ is then updated as a weighted sum of the top $k$ samples with the highest fitness
    \begin{equation}
    \boldsymbol{m}^{(g+1)}=\boldsymbol{m}^{(g)}+c_{\mathrm{m}} \sum_{i=1}^k w_i\left(\boldsymbol{x}_{i:n}^{(g+1)}-\boldsymbol{m}^{(g)}\right),
    \end{equation}
    where $\boldsymbol{x}_{i:n}^{(g+1)}$ denotes the $i$-th ranked sample, $w_i$ is the respective weight, and $c_{\mathrm{m}}$ is a learning rate, typically set to 1.
    \item The covariance matrix $\boldsymbol{C}^{(g)}$ is updated to adapt the shape of the distribution 
    \begin{eqnarray}
    \boldsymbol{C}^{(g+1)} &=& (1 - c_1 -c_k\sum_{i=1}^{n} w_i) \boldsymbol{C}^{(g)} \nonumber \\
    & + &c_1 ~\boldsymbol{p}_c^{(g+1)}{\boldsymbol{p}_c^{(g+1)}}^{\top} \nonumber \\
    & + &c_k \sum_{i=1}^{n} w_i \boldsymbol{y}_{i:n}^{(g+1)} {\boldsymbol{y}_{i:n}^{(g+1)}}^\top,
    \end{eqnarray}
    where $c_1$ and $c_k$ are learning rates for the covariance matrix adaptation, $\boldsymbol{y}_{i:n}^{(g+1)}=\left(\boldsymbol{x}_{i:n}^{(g+1)}-\boldsymbol{m}^{(g)}\right)/\sigma^{(g)}$, and $\boldsymbol{p}_c^{(g+1)}$ is the evolution path given by 
    \begin{eqnarray}
    \label{eq:evolution_path}
    \mathbf{p}_c^{(g+1)} \hspace{-2mm} &=& \hspace{-2mm} (1 - c_c)\:\mathbf{p}_c^{(g)} \nonumber \\ & +& \hspace{-2mm} \sqrt{c_c(2 - c_c)\mu_{\text{eff}}}\:\frac{\mathbf{m}^{(g+1)} - \mathbf{m}^{(g)}}{c_m\sigma^{(g)}},
    \end{eqnarray}
    with $\mu_{\text{eff}} = 1/\sum_{i=1}^{k} w_i^2,$ and $c_c$ the decay rate for the cumulation of the path.
    \item The step size $\sigma^{(g)}$
    is adjusted based on the evolution path $\boldsymbol{p}_\sigma^{(g+1)}$ to control the overall scale of the search
    \begin{eqnarray}
    \sigma^{(g+1)}=\sigma^{(g)} \exp \left(\frac{c_\sigma}{d_\sigma}\left(\frac{\left\|\boldsymbol{p}_\sigma^{(g+1)}\right\|}{\mathrm{E}\|\mathcal{N}(\mathbf{0}, \mathbf{I})\|}-1\right)\right),
    \end{eqnarray}
    where $c_\sigma$ and $d_\sigma$ are learning rates for the step-size adaptation, and $\mathrm{E}\|\mathcal{N}(\mathbf{0}, \mathbf{I})\|$ represents the expected length of a random vector sampled from a standard normal distribution.
\end{enumerate}
This process is repeated iteratively until a stopping criterion is met, such as reaching the maximum number of generations, achieving a target fitness value, or observing stagnation in the optimization progress. Please refer to Ref.~\cite{hansen2016cma} for a detailed description of this algorithm. 

In this work, since there are seven design parameters, the initial population size is 9 and the initial step size $\sigma_0$ is set to 0.16. We performed this algorithm using a Python implementation~\cite{hansen2019pycma} and we use its default setting for the rest of the parameters.

\subsection{Seed-pinning via steered molecular dynamics}\label{subsec:seed-pinning}
In this section, we present the enhanced sampling methods used to accelerate nucleation and growth of the zeolite in the MD simulations. Nucleation is typically a rare event within affordable timescales in MD simulations. To better capture nucleation events within limited simulation times, various enhanced sampling methods have been developed, e.g. umbrella sampling~\cite{torrie1977nonphysical}, forward flux sampling~\cite{allen2009forward}, seeding~\cite{bai_calculation_2006, espinosa2016seeding}, and variational umbrella seeding~\cite{gispen2024variational}, among others. The seeding approach involves inserting a small seed or crystalline cluster into the fluid medium. This crystal seed helps to overcome the high free-energy barrier associated with nucleation. This method is straightforward to implement and can be used to calculate nucleation rates in accordance with classical nucleation theory~\cite{farkas1927keimbildungsgeschwindigkeit, volmer1926keimbildung, becker1935annal}. However, in an inverse design setting, where the simulation parameters are initially far from optimal for zeolite self-assembly, the seed will often melt in the vast majority of simulations, providing little information about the fitness. To overcome this challenge, we propose a variant of the seeding approach called the seed-pinning method. Similar to the traditional seeding approach, a crystal seed is placed in the amorphous phase. The difference is that, similar to Ref.~\cite{sharma_nucleus-size_2018}, a restraining potential is applied to the seed size to prevent the seed from melting while allowing growth. This method of allowing only seed growth significantly reduces the nucleation barrier. Samples with favorable design parameters will continue to grow, while those with unfavorable design parameters will fluctuate around the restrained nucleus size. By monitoring these fluctuations the fitness of the sample can be evaluated. In the following, we will discuss how to measure these fluctuations and apply the restraint to the seed size.

 Due to the intricate structure of zeolite frameworks, traditional methods like Steinhardt bond-orientational order parameters~\cite{steinhardt_bond-orientational_1983} are insufficient for distinguishing zeolite-like T particles from amorphous T particles. Some level of distinction can be achieved by calculating bond-orientational order parameters for T particles with specific coordination numbers~\cite{kumar_two-step_2018}, but this is not straightforward to generalize across all framework types. To solve this problem, we employ an order parameter called environment similarity (ES)~\cite{piaggi_calculation_2019} to identify zeolite-like T particles. To evaluate the ES, an environment is defined as the closest $n$ neighbors of a given particle. ES compares the current environment, $\chi$, around a particle $i$ to a reference environment, $\chi_0$, using a normalized kernel $\tilde{k}$ defined as
\begin{equation}
    \tilde{k}_{\chi_0}(\chi) = \frac{1}{n} \sum\limits_{i\in\chi} \sum\limits_{j\in\chi_0} \exp\left( - \frac{|\boldsymbol{r}_i-\boldsymbol{r}^0_j|^2} {4\sigma^2} \right),
\end{equation}
where $\boldsymbol{r}_i$ are the coordinates of the particles, $\sigma$ is a broadening parameter to account for thermal fluctuations, and $n$ is the number of particles in the environment $\chi$. By definition, $\tilde{k}_{\chi_0}(\chi_0)=1$. In general, a particle can have multiple distinct environments in a crystal of a specific lattice, denoted as $\chi_1, \chi_2, \dots, \chi_{N_E}$, where $N_E$ denotes the number of possible environments. In this case, we adopt a best-match strategy to determine the environment, with the kernel defined as the maximum similarity among the $N_E$ reference environments 
\begin{equation}
    \tilde{k}_X(\chi)=\frac{1}{\lambda} \log \left(\sum_{l=1}^{N_E} \exp \left(\eta \tilde{k}_{\chi_l}(\chi)\right)\right).
\end{equation}
The largest value of $\tilde{k}_{\chi_l}(\chi)$ among the environments $\chi_{l} \in X$ is selected as the ES when $\eta$ is sufficiently large. Therefore, the ES order parameter ranges between 0 and 1. 

The data for the unique environments is extracted from the IZA~\cite{iza} and CDCC databases~\cite{ccdc} using the Environment Finder tool~\cite{github-environment-finder}. In general, we set a cutoff of 10 \AA~to search for unique environments, which is sufficient to capture the structural features needed for accurate zeolite identification. The number of distinct environments $N_E$ are listed in Table~\ref{tab:framework_setup} and vary from 6 for SOD to 136 for Z1. The number of particles $n$ in an environment typically varies from 55 to 75. Considering the notable fluctuations at high temperatures, the value of $\sigma$ is set to 0.5 \AA.

It is worth noting that this order parameter is not rotationally invariant. This means that even if the identified particles have the same structure as the reference environment, a high ES value will not be obtained if the orientation is inconsistent. As a result, our protocol is limited to zeolites that are aligned with the chosen references. However, we address this by starting with an aligned seed and applying restraints to it to preserve that alignment. To further restrain the seed size, we bias the number of particles $N'$ with an ES order parameter greater than a threshold, $\tilde{k}_{\chi}(\chi) > k'$, using a rational switching function to ensure that $N'$ is continuous and differentiable.

First, we insert a seed composed solely of T particles, extracted from the target zeolite framework, into the amorphous mixture system. For all cases, a seed is cut to include as many T-T bonds as possible while keeping the pores open. We then apply a harmonic potential to the seed particles to restrain the ES order parameter~\cite{torrie1977nonphysical}
\begin{equation}
    \beta V(t) = \frac{1}{2} \kappa \left(N'_\text{seed}(t)- N_\text{seed}\right)^2,
    \label{eq:harmonic}
\end{equation}
where $\kappa$ is a dimensionless constant, $N_\text{seed}$ is the initial number of T particles in the seed, and $N'_\text{seed}$ is the number of T particles from the original seed that retains an ES order parameter above a given threshold, $k'$. This bias is only applied to the seed particles that originally formed the seed, which has two advantages: first, the T particles that are not originally in the seed are free to either remain in the amorphous phase or nucleate, and second, the computational cost for calculating the ES and its gradient is invested only on a reduced number of T particles. For the fitness calculation, the ES is evaluated for all particles, $N'_\text{system}$, but this quantity does not need to be computed at every timestep. While it is in principle possible to calculate the nucleation free energy $\Delta G$ using seed-pinning, our goal is to enhance sampling around the nucleation barrier and use the fluctuations to assess the fitness for a given set of design parameters. Once a high fitness is achieved, the optimal design parameters are tested in a self-assembly simulation without any seed or bias. 

\subsection{Simulation protocol}

\begin{table*}
    \centering
    \setlength{\tabcolsep}{2pt}
    \begin{tabular}{ccccccccccccccc}
    \hline
        Framework & $N_U$ & $N_E$ & $N_S$ & $N_{seed}$ & $\kappa$ & $\chi_{\mathrm{T}}$ & $T$ (K) & $\varepsilon_{\mathrm{TT}}$ (eV) & $\varepsilon_{\mathrm{TS}}$ (eV) & $\varepsilon_{\mathrm{SS}}$ (eV) & $\sigma_{\mathrm{TS}}$ (\AA) & $\sigma_{\mathrm{SS}}$ (\AA) & $\lambda$ \\\hline
        Z1  & 132 & 136* & 1056 & 50 & 5 & 0.74 & [600, 700] & [0.3, 0.7] & [0.05, 0.09] & [0.02, 0.09] & [4, 7] & [4, 8] & [15, 25] \\
        SGT & 64 & 32 & 1024 & 23 & 20 & 0.94 & [600, 700] & [0.3, 0.7] & [0.05, 0.09] & [0.02, 0.09] & [4, 7] & [4, 8] & [15, 25] \\
        SOD & 12 & 6 & 768 & 23 & 20 & 0.86 & [600, 700] & [0.3, 0.7] & [0.05, 0.09] & [0.02, 0.09] & [4, 7] & [6, 10] & [15, 25] \\
        AFI & 24 & 24 & 1920 & 192 & 10 & 0.92 & [600, 700] & [0.3, 0.7] & [0.05, 0.09] & [0.02, 0.09] & [4, 7] & [4, 9] & [15, 25] \\
        CFI & 32 & 16 & 2048 & 256 & 5 & 0.94 & [600, 700] & [0.3, 0.7] & [0.05, 0.09] & [0.01, 0.05] & [4, 7] & [4, 9] & [15, 25] \\
    \hline
    \end{tabular}
    \caption{The simulation set-up and design parameter boundaries for each framework. Here, $N_U$ represents the number of T particles in a unit cell, $N_E$ denotes the number of unique environments, and $N_S$ is the total number of T particles within the seed-pinning simulation box. $N_\text{seed}$ indicates the seed size, $\kappa$ denotes the dimensionless constant used in Eq.~\ref{eq:harmonic}, $T$ the temperature, $\varepsilon_{\mathrm{TT}}$, $\varepsilon_{\mathrm{TS}}$, and $\varepsilon_{\mathrm{SS}}$ the interaction strength between TT, TS, and SS particle pairs, respectively, and $\sigma_{\mathrm{TS}}$ and $\sigma_{\mathrm{SS}}$ the interaction scale between TS and SS particle species, respectively. *: The number of environments for Z1 is determined from scaled FIR-30 unit cells, which contain a larger number of particles.}
    \label{tab:framework_setup}
\end{table*}

In this study, we categorize all zeolite frameworks into three types based on channel dimensionality: cage-type (channel dimensionality = 0), where S particles are confined within cages and cannot move freely; hole-type (channel dimensionality $\ge$ 1), where S particles are mobile; and framework-type, which includes both cages and holes (channels). This workflow was first applied to reproduce a framework-type Z1, and a cage-type SGT~\cite{sgt} zeolite, as documented in previous studies~\cite{kumar_two-step_2018, dhabal_coarse-grained_2021, wang_homochiral_2015, kumar_could_2018}. The Z1 zeolite structure is based on the structure of the MOF FIR-30~\cite{FIR-30}, see for more details Section IIIA. Subsequently, we focused on reproducing a new cage-type SOD~\cite{sod}, and a new hole-type CFI~\cite{cfi} obtained from the IZA database.

For cage-type and framework-type zeolites, the number of S particles per cage is estimated based on geometric constraints, typically assuming one S particle per cage. A supercell is then constructed for each zeolite using unit cells from the IZA database, typically containing about a thousand T particles to balance computational efficiency and finite-size effects. Next, a seed is extracted from the supercell. The seed is selected according to two key principles: first, it must remain open and not form a closed cage, allowing the passage of S and T particles; second, the shape and size of the seed are tailored to the target zeolite framework. Table~\ref{tab:framework_setup} presents the specific settings for each framework. While larger seeds may promote crystal growth in MD simulations, i.e. without bias, crystallization often does not occur in self-assembly tests using the same design parameters due to a high nucleation barrier $\Delta G$. Therefore, the seed should be kept as small as possible while still effectively inducing nucleation. The seed is then embedded in a simulation box with dimensions identical to the supercell, and with additional T and S particles randomly placed according to the target composition. Subsequently, a one-nanosecond MD simulation is performed in the isobaric-isothermal ($NPT$) ensemble at a pressure of $P=0$ bar and a time step of 5 fs. The pressure is controlled isotropically in all three Cartesian coordinates, with the masses of both T and S particles set to 28.0855 u, equivalent to the mass of silicon. The average environment similarity (ES) of the T particles from the last quarter of the simulation trajectory is used as the fitness value for the optimizer, which then initiates the next iteration.

After 50 rounds of MD simulations, high-fitness solutions are tested in self-assembly simulations using larger system sizes, without the use of any seeds or biases, to assess their effectiveness. In these tests, the pressure is controlled independently along each of the three Cartesian coordinates.

All simulations are conducted using LAMMPS (version 2Aug2023)~\cite{LAMMPS}, with biasing implemented through PLUMED (version 2.8.3)~\cite{bonomi2009plumed, tribello2014plumed}. Snapshots are rendered using the OVITO software~\cite{ovito}.

%% file: results_z1.tex
\subsection{Framework-type Z1 zeolite}

\begin{figure}
    \centering
    \includegraphics[width=1.0\linewidth]{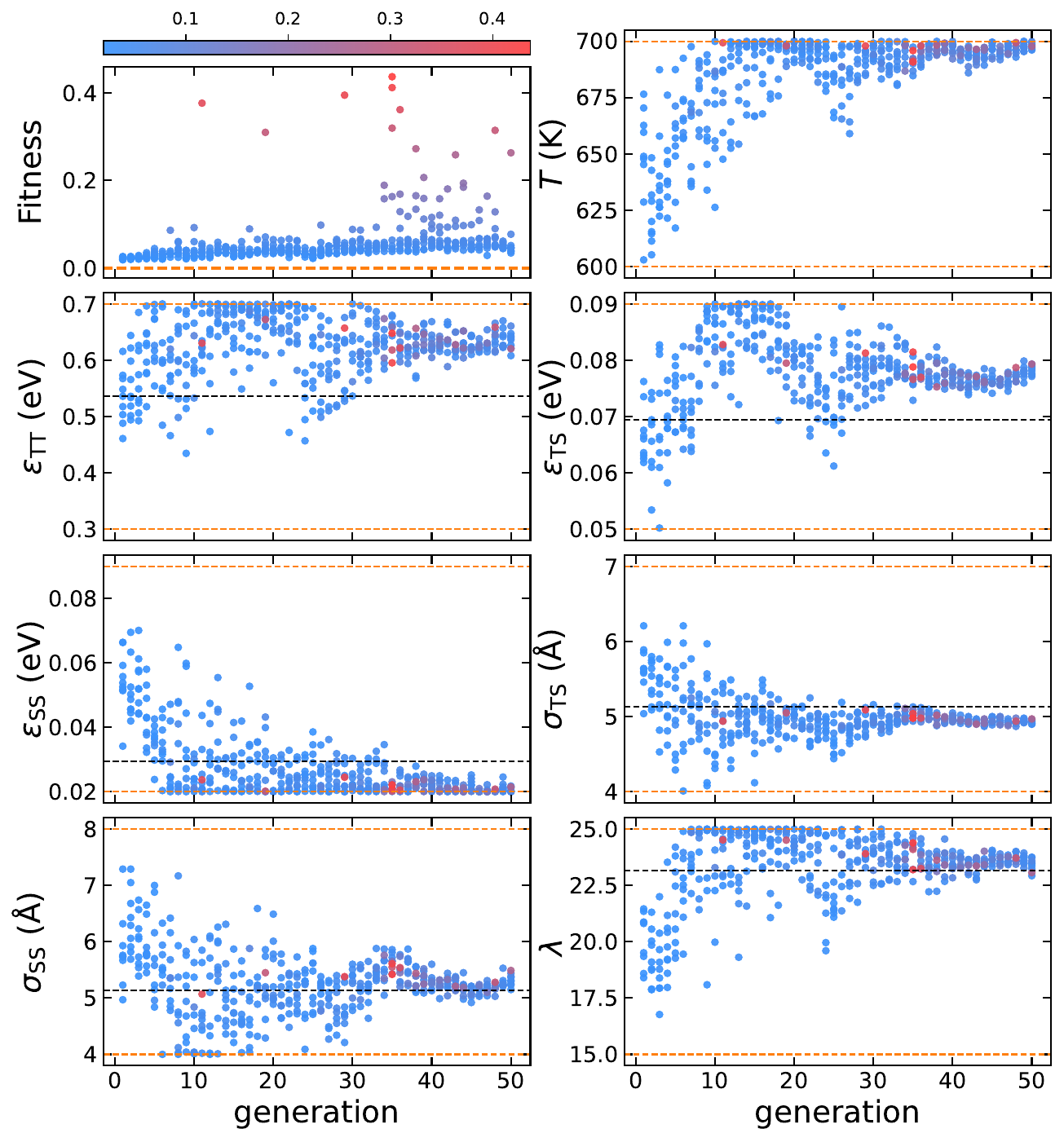}
    \caption{Evolution of the design parameters for Z1, with each point colored according to its fitness. The dashed orange lines represent the parameter boundaries, while the dashed black lines correspond to the values of Ref.~\cite{bertolazzo_unstable_2022}.}
    \label{fig:z1_evo_50}
\end{figure}

\label{sec:z1}
\begin{figure}
    \centering
    \includegraphics[width=1.0\linewidth]{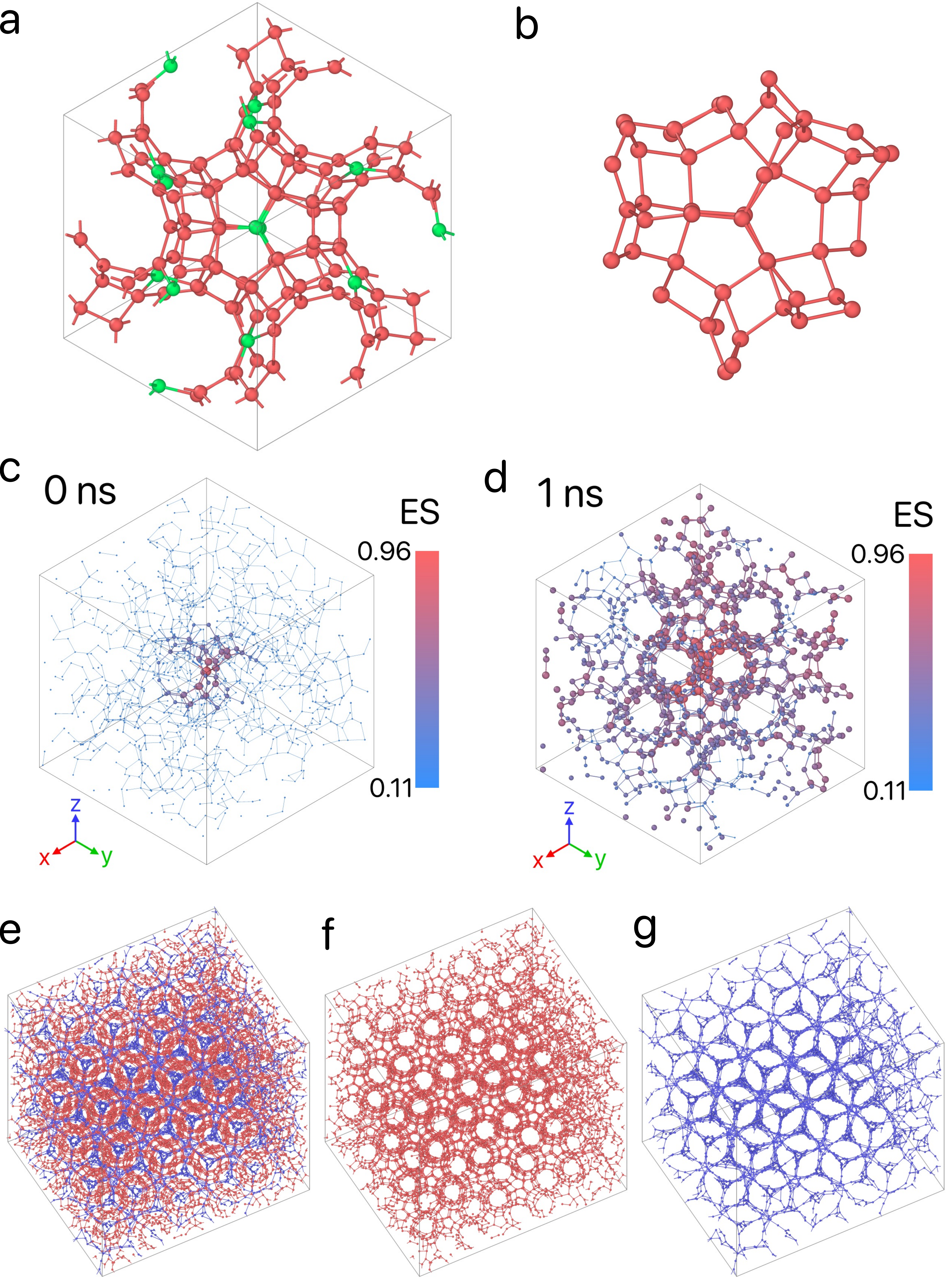}
    \caption{(a) The unit cell of the MOF FIR-30 framework~\cite{FIR-30}, consisting of 120 Zn atoms of FIR-30 (shown in red) and 16 additional particles (shown in green)~\cite{kumar_could_2018}. This FIR-30 unit cell, composed of 136 particles, is used to create a supercell, with coordinates scaled so that the neighbor distance of the Zn atoms is 3.3 \AA, matching the typical T-T interparticle distance. (b) A crystalline seed of 50 particles is extracted from this FIR-30 supercell. (c) The initial configuration at 0 ns and (d) the final configuration at 1 ns of a seed-pinning simulation for the highest-fitness sample from the 29th generation of the inverse design protocol for Z1. The radius of the shown T particles is proportional to their environment similarity (ES) to the target framework, as indicated by the colorbar. For clarity, S particles are not shown. (e), (f) and (g) The Z1 framework obtained from an unbiased self-assembly simulation using the highest-fitness solution from the 29th generation. (e) The intertwined structure of T and S particles. (f) The porous structure formed by T particles. (g) The gyroid network formed by S particles.}
    \label{fig:z1}
\end{figure}

We start our investigation by reverse-engineering the framework-type Z1 within a coarse-grained zeolite model consisting of a binary mixture of T and S particles. Our goal is to identify the optimal set of design parameters that facilitates the self-assembly of Z1  within this mixture using the protocol as outlined in Section II.

The zeolite Z1, as examined in Ref.~\cite{kumar_could_2018}, has a T particle structure, derived from the Zn atom positions in the unit cell of the MOF FIR-30~\cite{FIR-30}. We obtain the coordinates of the 120 Zn atoms in the unit cell of MOF FIR-30 from the supporting information of Ref.~\cite{wang2015homochiral}. These 120 Zn atoms are represented by the red particles in Fig.~\ref{fig:z1}(a). We then scale the coordinates so that the nearest neighbor distance between Zn atoms is set to 3.3 \AA, matching the typical T-T interparticle distance. To convert the 3-coordinated particles at the surface of the triangular channels to the 4-coordinated particles as observed in Z1, we add 16 additional T particles, represented by the green particles in Fig.~\ref{fig:z1}(a). This adjustment brings the total number of T particles in the FIR-30 unit cell to 136, closely resembling the Z1 structure, which contains 132 particles. Using this modified unit cell, we create a supercell of FIR-30, from which we extract a crystalline seed consisting of 50 particles as shown in Fig.~\ref{fig:z1}(b) for the seed-pinning simulations. Additionally, we identify the unique environments for evaluating the environment similarity order parameter. The total number of unique environments of Z1 is as large as 136 with typically 55-75 particles in each environment.

For our inverse design protocol, we consider the following seven design parameters, $\varepsilon_{\mathrm{TT}}$, $\varepsilon_{\mathrm{TS}}$, $\varepsilon_{\mathrm{SS}}$, $\sigma_{\mathrm{TS}}$, $\sigma_{\mathrm{SS}}$, $\lambda$, and temperature $T$. The values of these parameters are sampled at each generation from a multivariate Gaussian distribution according to the CMA-ES algorithm as outlined in Section IIC. For each set of parameters, also called sample, we run a short MD simulation of 1 ns, where a crystalline seed is immersed in a disordered fluid using the seed-pinning method. A typical example of the initial and final configurations of such a seed-pinning simulation is shown in Fig.~\ref{fig:z1}(c,d). The environment similarity order parameter is then used to evaluate the fitness of each sample. This information is used by the CMA-ES optimizer.

We run the evolution strategy for 50 generations, and present the results of this reverse-engineering process in Fig.~\ref{fig:z1_evo_50}. Although the average fitness value increases gradually over time, we identified several high-fitness samples that stand out throughout the process. Additionally, the evolution of the design parameters is shown in Fig.~\ref{fig:z1_evo_50}, indicating that the inverse design protocol converges toward a specific range of parameter values, where high-fitness values are found. For instance, in the high-fitness solutions, $\sigma_{\mathrm{TS}}$, $\sigma_{\mathrm{SS}}$, $\varepsilon_{\mathrm{SS}}$, and $\lambda$ converged to 5.0 \AA, 5.3 \AA, 0.022 eV, and 24.3, respectively---values close to the previously reported 5.13 \AA, 5.13 \AA, 0.029 eV, and 23.15, respectively~\cite{bertolazzo_unstable_2022}. However, the interaction strengths $\varepsilon_{\mathrm{TT}}$, $\varepsilon_{\mathrm{TS}}$, and the temperature $T$, converged to 0.653 eV, 0.081 eV, and 698 T, which are significantly higher than those reported in Ref.~\cite{bertolazzo_unstable_2022} indicated by the horizontal black dashed lines in Fig.~\ref{fig:z1_evo_50}.

To validate our inverse-design protocol, we perform unbiased self-assembly simulations using the top eight highest-fitness parameter combinations. Three of these parameter combinations, listed in Table~\ref{tab:z1_solutions}, spontaneously nucleate and grow into Z1, as shown in Fig.~\ref{fig:z1}(e)-(g). Compared to the parameters reported in Ref.~\cite{bertolazzo_unstable_2022}, our solutions exhibit larger $\varepsilon_{\mathrm{TT}}$ and $\varepsilon_{\mathrm{TS}}$, which explains their ability to self-assemble into Z1 at elevated temperatures. Additionally, the self-assembly simulations showed no mesophases during the nucleation pathway of Z1, aligning with a recently published rescaled zeolite model where the temperature is below 636 K~\cite{bertolazzo_unstable_2022}. Overall, our optimized design parameters closely match previous results, as shown in Table~\ref{tab:z1_solutions}.

\begin{table}
    \setlength{\tabcolsep}{3pt}
    \centering
    \begin{tabular}{cccccccc}
    \hline
        Solution & $T$ & $\varepsilon_{\mathrm{TT}}$ & $\varepsilon_{\mathrm{TS}}$ & $\varepsilon_{\mathrm{SS}}$ & $\sigma_{\mathrm{TS}}$ & $\sigma_{\mathrm{SS}}$ & $\lambda$ \\
        in     & K   & eV    & eV    & eV    & \AA  & \AA  &       \\\hline
        Gen 11 & 699 & 0.630 & 0.083 & 0.023 & 4.94 & 5.07 & 24.52 \\
        Gen 19 & 698 & 0.672 & 0.080 & 0.020 & 5.05 & 5.45 & 24.51 \\
        Gen 29 & 698 & 0.657 & 0.081 & 0.024 & 5.09 & 5.38 & 23.91 \\
        Ref.~\cite{bertolazzo_unstable_2022}& $<$654 & 0.537 & 0.069 & 0.029 & 5.13 & 5.13 & 23.15 \\
    \hline
    \end{tabular}
    \caption{The high-fitness solutions that spontaneously self-assemble into the framework-type Z1 zeolite within 50 ns.}
    \label{tab:z1_solutions}
\end{table}

\FloatBarrier

%% file: results_sgt.tex
\subsection{Cage-type SGT zeolite}

\begin{figure}
    \centering
    \includegraphics[width=1.0\linewidth]{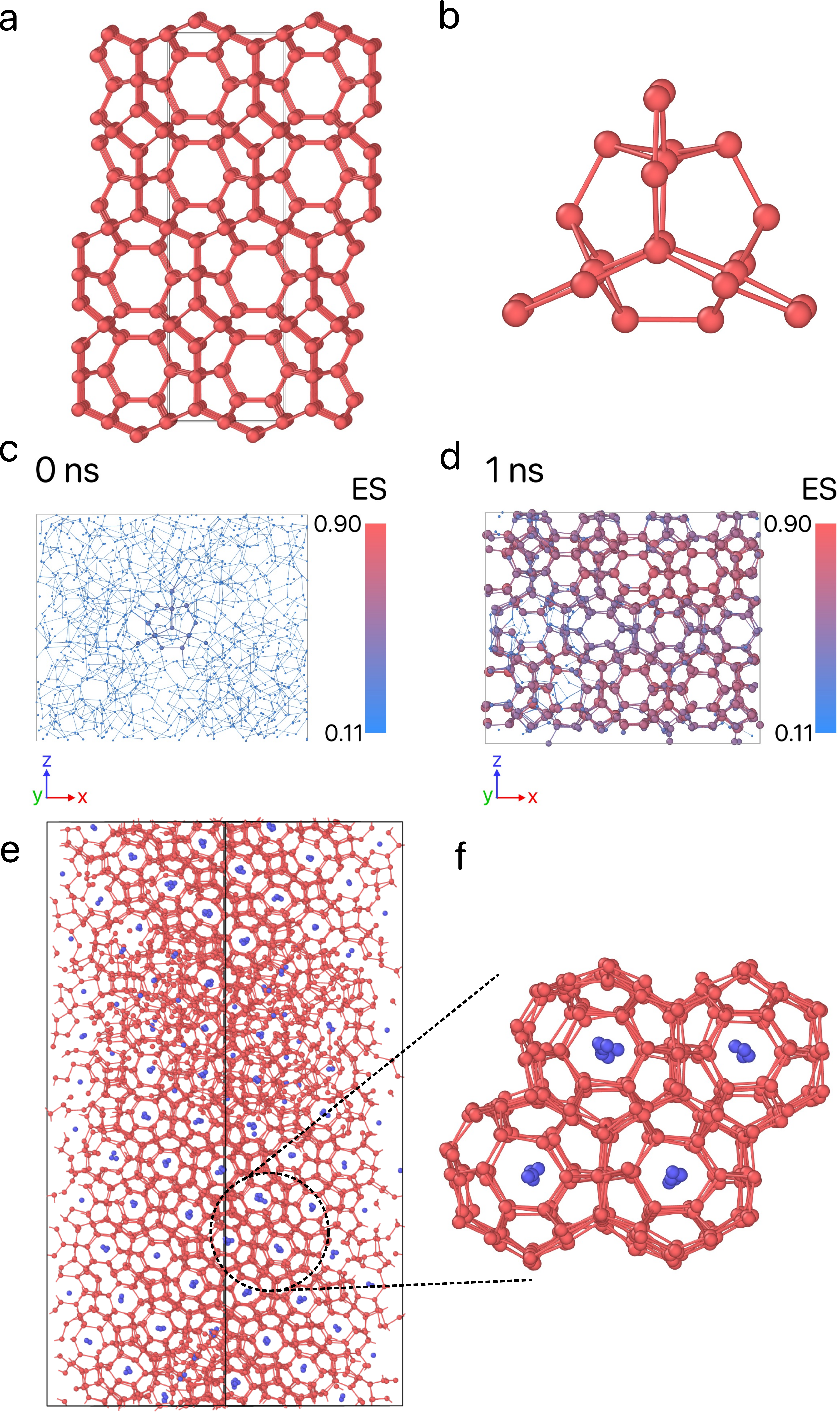}
    \caption{(a) The structure of the perfect SGT framework~\cite{sgt}. (b) A crystalline seed composed of 23 T particles. (c) The initial configuration is at 0 ns, and (d) the final configuration is at 1 ns of a seed-pinning simulation for the highest-fitness sample from the 44th generation of the inverse design protocol for SGT. The radius of the shown T particles is proportional to their environment similarity (ES) to the target framework, as indicated by the colorbar. For clarity, S particles are not shown. (e) The SGT framework obtained from an unbiased self-assembly simulation using the highest-fitness solution from the 44th generation. This configuration shows an AB random stacking structure.}
    \label{fig:sgt}
\end{figure}

\begin{figure}[H]
    \centering
    \includegraphics[width=1.0\linewidth]{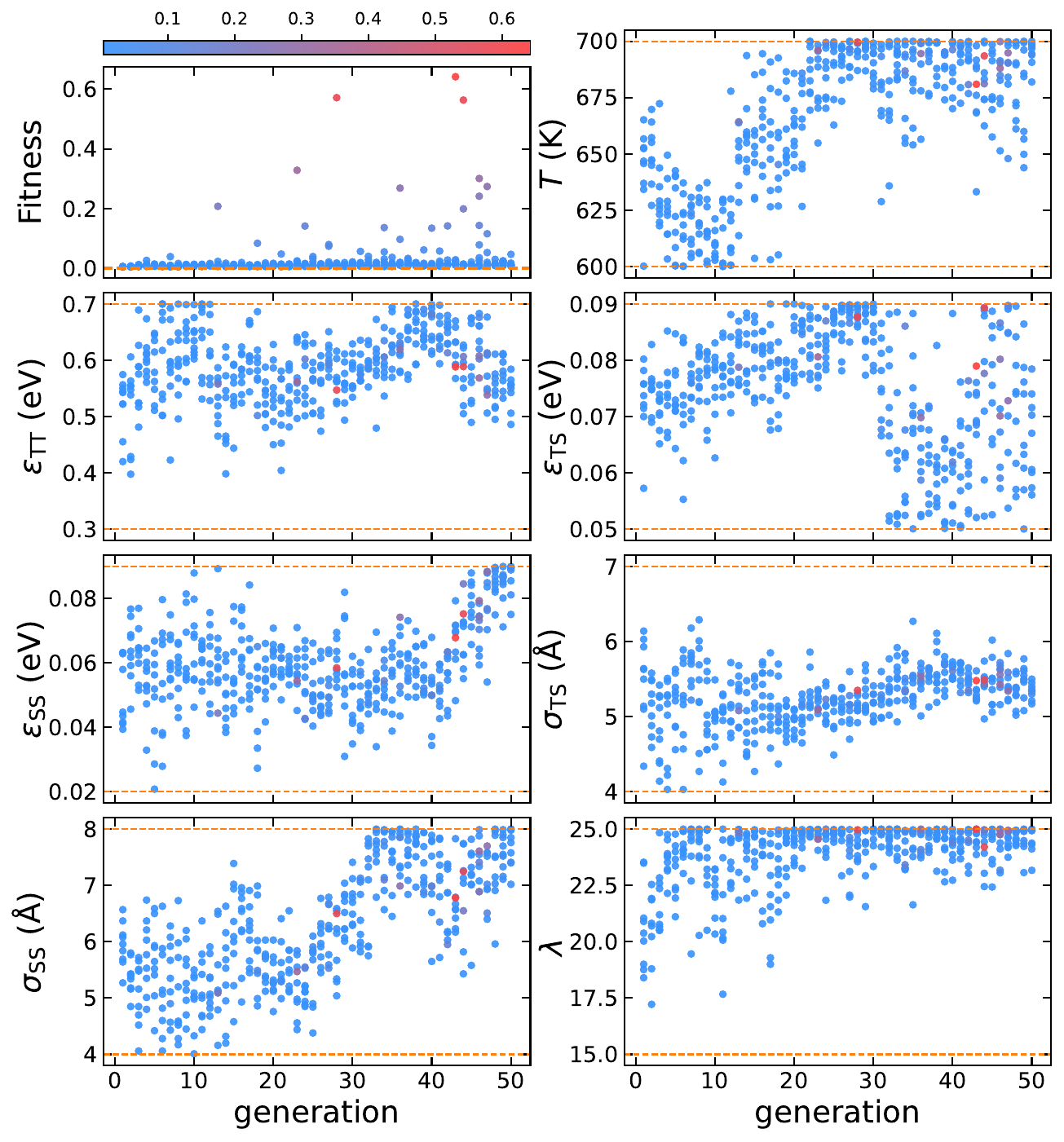}
    \caption{Evolution of the design parameters for SGT, with each point colored according to its fitness. The dashed orange lines represent the parameter boundaries.}
    \label{fig:sgt_evo_23}
\end{figure}

We now use our inverse design protocol to find the optimal set of design parameter values that facilitate the self-assembly of the cage-type SGT zeolite within the coarse-grained zeolite model of T and S particles. SGT is a typical high-purity silica zeolite with a predominantly tetrahedral structure and, according to our classification method, falls under the cage-type zeolites. We obtain the coordinates of the T particles for SGT from the IZA database~\cite{sgt}, and present the structure in Fig.~\ref{fig:sgt}(a). We assume that each large cage contains one S particle. Given that the SGT unit cell consists of 64 T particles and 4 large cages, the composition $\chi_{\mathrm{T}}$ is set at $64/(64+4) \approx 0.94$. For the inverse design process, we use a small seed of 20 T particles, as shown in Fig.~\ref{fig:sgt}(b), chosen as the minimal size necessary to induce nucleation. In addition, we identify the unique environments for evaluating the environment similarity order parameter. The number of unique environments for SGT is 32 as shown in Table.~\ref{tab:framework_setup}. We follow the same inverse design protocol as described in Section ~\ref{sec:z1} for the inverse design of Z1, and consider again seven design parameters, $\varepsilon_{\mathrm{TT}}$, $\varepsilon_{\mathrm{TS}}$, $\varepsilon_{\mathrm{SS}}$, $\sigma_{\mathrm{TS}}$, $\sigma_{\mathrm{SS}}$, $\lambda$, and $T$, sampled from a multivariate Gaussian distribution according to the CMA-ES algorithm. For each set of parameters, we run a short MD simulation of 1 ns using the seed-pinning method. Fig.~\ref{fig:sgt}(c,d) shows the initial and final configurations for the parameter combination corresponding to generation 44 of a seed-pinning simulation. The fitness of each sample is evaluated using the environment similarity order parameter, and these fitness values are then used by the CMA-ES optimizer. We run the evolution strategy for 50 generations, with the results for the evolution of the design parameters presented in Fig.~\ref{fig:sgt_evo_23}.

The evolution of the design parameters reveals high-fitness solutions across a broader range of values compared to Z1, suggesting that the self-assembly conditions for SGT are less restrictive. Additionally, we observe that the inverse design protocol converges for only a few design parameters, within a specific range where high fitness values are found. For example, $T$, $\sigma_{\mathrm{TS}}$, and $\lambda$ converge to 686 K, 5.4 \AA, and 24.7, respectively, within the high-fitness solutions, while $\varepsilon_{\mathrm{TT}}$, $\varepsilon_{\mathrm{TS}}$, $\varepsilon_{\mathrm{SS}}$, and $\sigma_{\mathrm{SS}}$ exhibit considerable variability.

To validate the wide variability in the parameter values of the high-fitness solutions, we run unbiased self-assembly simulations using the top ten highest-fitness solutions. We identify six solutions that spontaneously crystallize into SGT within 50 ns. We present the corresponding parameter values in Table~\ref{tab:sgt_solutions}. Previously reported parameters that facilitated SGT self-assembly match those of Z1, except for a higher $\chi_{\mathrm{T}}$~\cite{kumar_could_2018}. Note that these parameters are based on the T-S model without rescaling the T-T distance~\cite{bertolazzo_unstable_2022}.

Our optimized values for $\varepsilon_{\mathrm{TT}}$ and $\lambda$ in the high-fitness solutions closely match previous results~\cite{kumar_could_2018}, which is expected, as SGT is primarily a pure silica zeolite with strong tetrahedral interactions between T particles. However, the values for $\varepsilon_{\mathrm{SS}}$ and $\sigma_{\mathrm{SS}}$ differ significantly. The optimized $\sigma_{\mathrm{SS}}$ values range from 5.09 to 7.61 \AA, accompanied by a gradual increase in $\varepsilon_{\mathrm{SS}}$ from 0.044 to 0.079 eV. The larger interaction distance, $\sigma_{\mathrm{SS}}$, necessitates a stronger interaction strength, $\varepsilon_{\mathrm{SS}}$, to achieve an effective shaping effect. This indicates that rather than narrow optimization valleys, there are broader channels in the optimization space, where trade-offs between parameters allow various types of S particles to guide the formation of the same T particle framework.
\begin{table}[H]
    \setlength{\tabcolsep}{3pt}
    \centering
    \begin{tabular}{cccccccc}
    \hline
        Solution & $T$ & $\varepsilon_{\mathrm{TT}}$ & $\varepsilon_{\mathrm{TS}}$ & $\varepsilon_{\mathrm{SS}}$ & $\sigma_{\mathrm{TS}}$ & $\sigma_{\mathrm{SS}}$ & $\lambda$ \\
        in     & K   & eV    & eV    & eV    & \AA  & \AA  &       \\\hline
        Gen 13 & 664 & 0.558 & 0.079 & 0.044 & 5.08 & 5.09 & 24.85 \\
        Gen 23 & 696 & 0.562 & 0.081 & 0.055 & 5.09 & 5.47 & 24.55 \\
        Gen 36 & 695 & 0.619 & 0.070 & 0.074 & 5.54 & 6.99 & 25.00 \\
        Gen 43 & 681 & 0.588 & 0.079 & 0.068 & 5.48 & 6.78 & 25.00 \\
        Gen 44 & 694 & 0.589 & 0.089 & 0.075 & 5.49 & 7.25 & 24.21 \\
        Gen 46 & 688 & 0.568 & 0.070 & 0.079 & 5.43 & 7.61 & 24.76 \\
        Ref.~\cite{kumar_could_2018} &     & 0.537 & 0.069 & 0.029 & 5.13 & 5.13 & 23.15 \\
    \hline
    \end{tabular}
    \caption{The high-fitness solutions that spontaneously self-assemble into the cage-type SGT zeolite within 50 ns.}
    \label{tab:sgt_solutions}
\end{table}

%% file: results_sod.tex
\subsection{Cage-type SOD zeolite and by-products}

\begin{figure}
    \centering
    \includegraphics[width=1.0\linewidth]{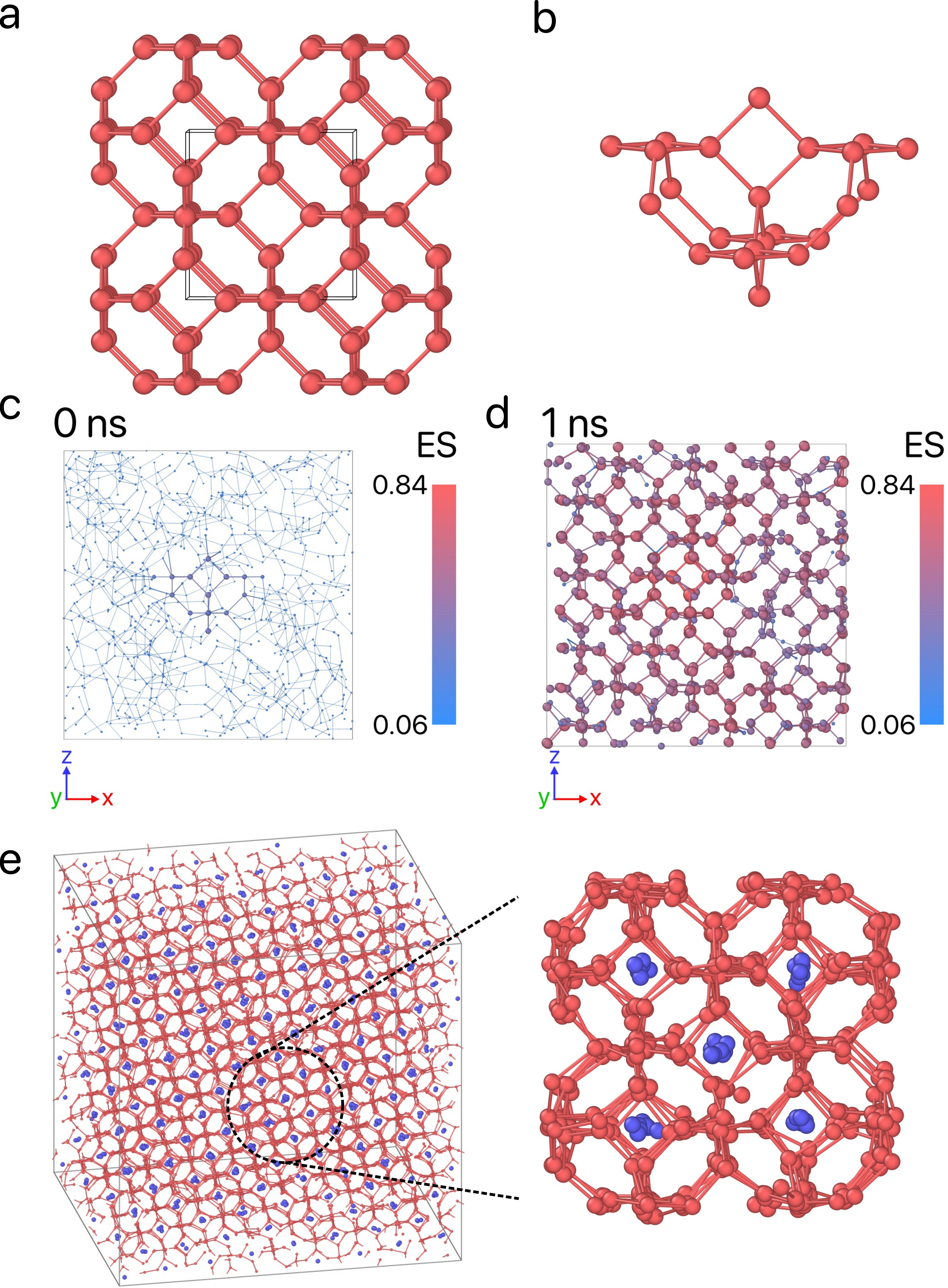}
    \caption{(a) The structure of the perfect SOD framework~\cite{sod}. The black box indicates the unit cell. (b) A crystalline seed composed of 23 T particles. (c) The initial configuration at 0 ns and (d) the final configuration at 1 ns of a seed-pinning simulation for the highest-fitness sample from the 40th generation of the inverse design protocol for SOD. The radius of the shown T particles is proportional to their environment similarity (ES) to the target framework, as indicated by the colorbar. For clarity, S particles are not shown. (e) The SOD framework obtained from an unbiased self-assembly simulation using the high-fitness solution from the 40th generation. To our knowledge, there are no previously reported parameters for SOD in the T-S model.}
    \label{fig:sod}
\end{figure}

After the successful reproduction of Z1 and SGT, our inverse design protocol has proven its effectiveness in targeting frameworks known to self-assemble within the T-S model. We now aim to design frameworks that have not yet been reproduced using the T-S model. We start with SOD, one of the simplest zeolite frameworks, characterized by only six unique environments. The SOD unit cell contains 12 T particles and includes 2 cages, resulting in a T particle composition of $\chi_{\mathrm{T}}=12/(12+2)\approx0.86$. The perfect SOD crystal and the crystalline seed used for seed-pinning are shown in Figs.~\ref{fig:sod}(a,b), respectively. 

\begin{figure}[H]
    \centering
    \includegraphics[width=1.0\linewidth]{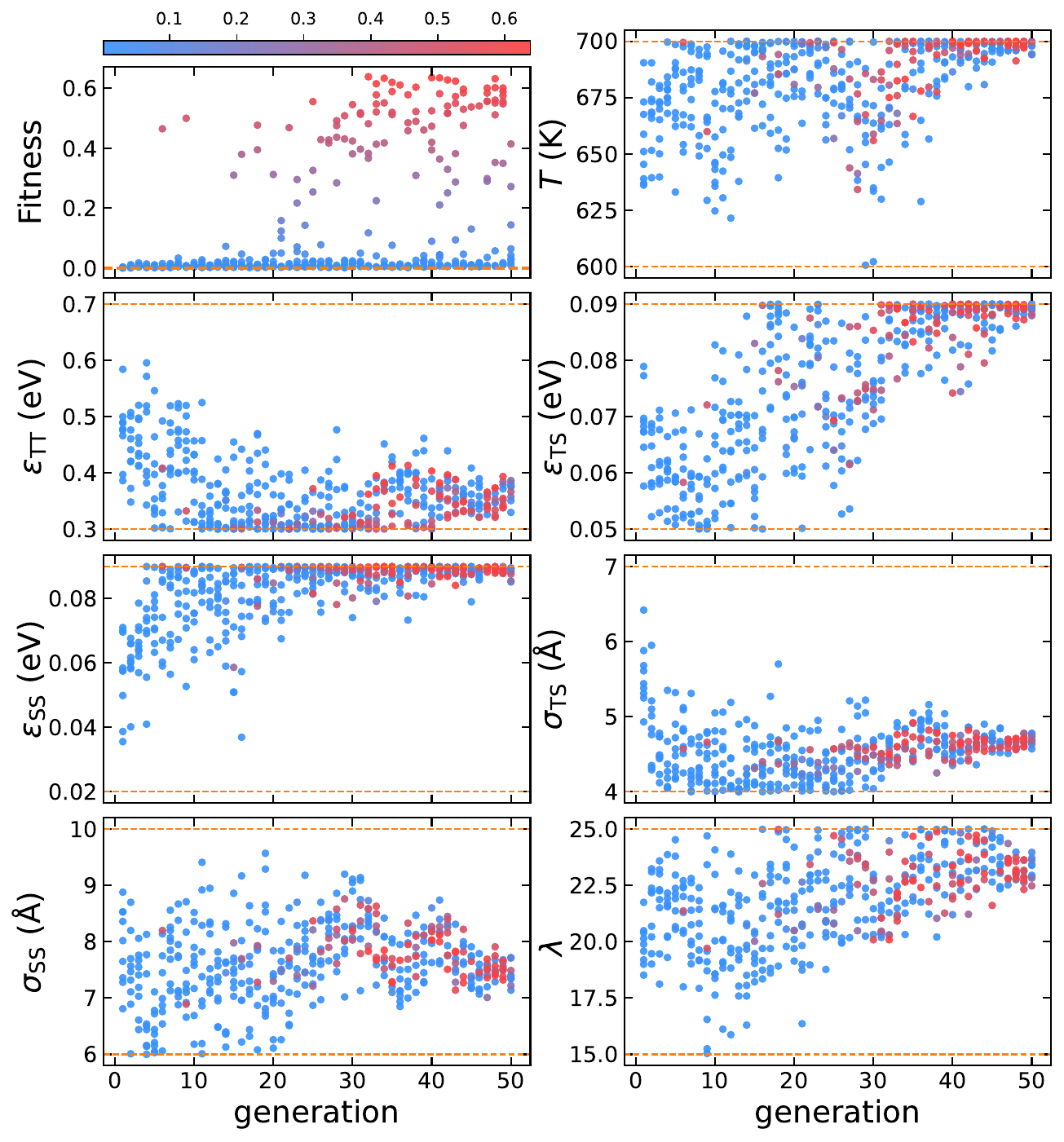}
    \caption{Evolution of the design parameters for SOD, with each point colored according to its fitness. The dashed orange lines represent the parameter boundaries.}
    \label{fig:sod_evo_19}
\end{figure}

In Fig.~\ref{fig:sod_evo_19}, we present the evolution of the fitness values during the optimization process for SOD self-assembly. We use a seed of size 23 T particles. Similar to the trends observed in the evolution of Z1 and SGT, a consistent high fitness across all samples within a generation is not achieved due to the rare event nature of nucleation. However, from generation 25 onward, a significant number of samples per generation exhibits a high fitness. In comparison to the optimized parameters for Z1 and SGT, the value of $\varepsilon_{\mathrm{TT}}$ decreases, indicating that a weaker tetrahedral interaction strength facilitates the formation of 4-membered and 6-membered rings in SOD~\cite{noya_assembly_2019}. Additionally, $\sigma_{\mathrm{SS}}$ converges to approximately $8$ \AA, which is close to the distance between S particles in the SOD framework. Two high-fitness samples (fitness $\geq$ 0.15) were obtained in the 40th and 41st generations, as shown in Table~\ref{tab:sod_solutions}. Both of these samples show very similar parameter values and successfully produced a SOD framework in an unbiased self-assembly simulation. Fig.~\ref{fig:sod}(c) shows the nucleated SOD framework using the parameters from the 40th generation. When comparing these optimized SOD parameters to those for ZI and SGT, we observe significant differences in $\varepsilon_{\mathrm{TT}}$ across the different frameworks. This observation aligns with the literature, which indicates that the selection of framework is primarily governed by the interaction between T particles, with $\varepsilon_{\mathrm{TT}}$ playing a key role in determining the phase behavior of this model~\cite{kumar_assembly_2019}. Similarly, $\sigma_{\mathrm{TS}}$ is optimized to geometrically support the target framework, generally matching the radius of the inscribed sphere in the SOD cage when a single S particle is present.

\begin{table}[H]
    \setlength{\tabcolsep}{3pt}
    \centering
    \begin{tabular}{cccccccc}
    \hline
        Solution & $T$ & $\varepsilon_{\mathrm{TT}}$ & $\varepsilon_{\mathrm{TS}}$ & $\varepsilon_{\mathrm{SS}}$ & $\sigma_{\mathrm{TS}}$ & $\sigma_{\mathrm{SS}}$ & $\lambda$ \\
        in     & K   & eV    & eV    & eV    & \AA  & \AA  &       \\\hline
        Gen 40 & 700 & 0.403 & 0.090 & 0.089 & 4.75 & 8.03 & 23.36 \\
        Gen 41 & 698 & 0.382 & 0.089 & 0.087 & 4.66 & 8.12 & 22.65 \\
    \hline
    \end{tabular}
    \caption{The high-fitness solutions that spontaneously self-assemble into the cage-type SOD zeolite within 50 ns.}
    \label{tab:sod_solutions}
\end{table}

\begin{figure}[H]
    \centering
    \includegraphics[width=1.0\linewidth]{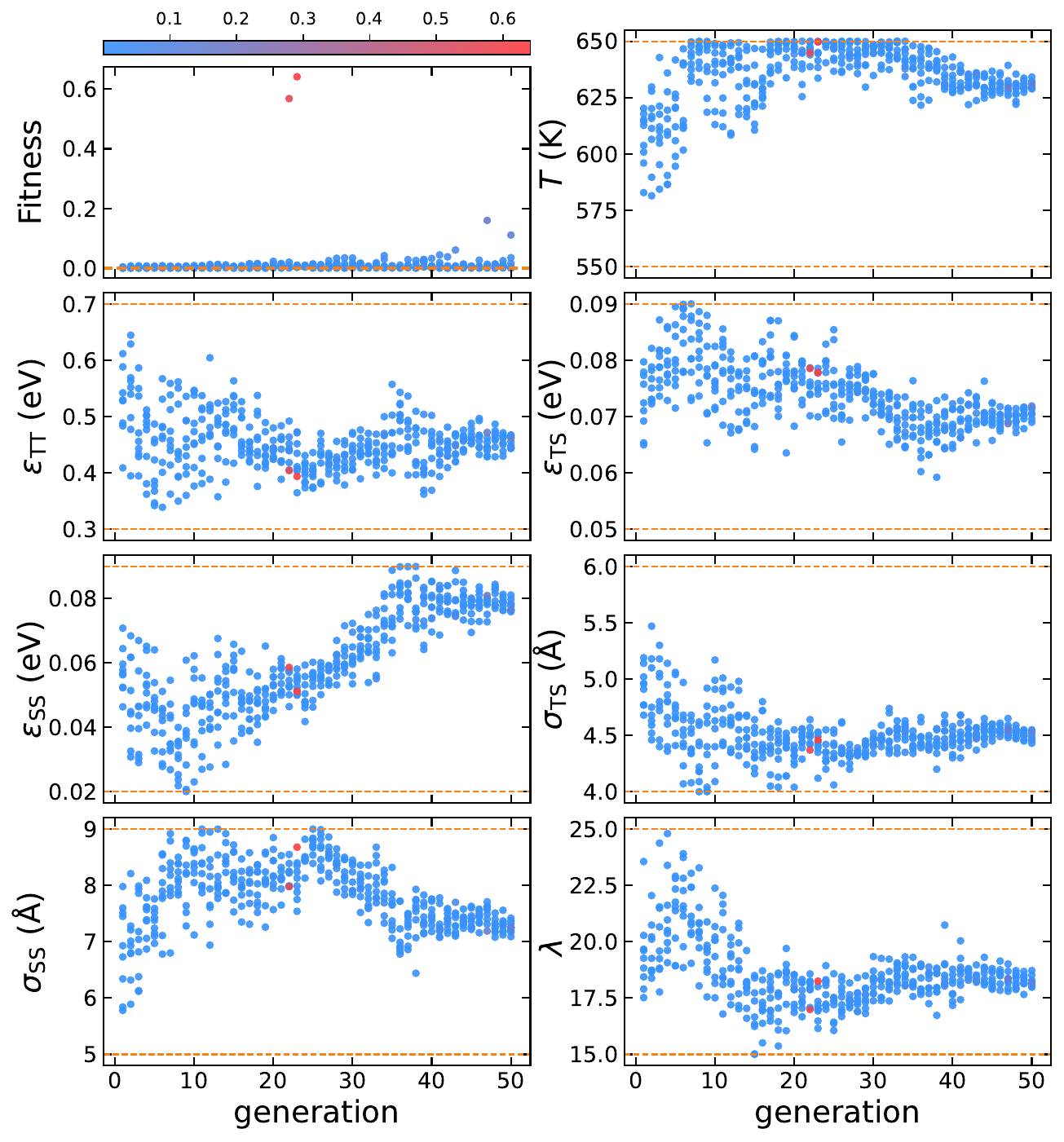}
    \caption{Evolution of the design parameters for SOD with a crystalline seed of 19 T particles and a lower temperature range. All the points are colored according to their fitness. The dashed orange lines represent the parameter boundaries.}
    \label{fig:sI_evo_19}
\end{figure}

During the inverse design of SOD, when using a seed of 19 rather than 23 T particles, we unexpectedly discovered the formation of the sI clathrate---also known by the zeolite code name MEP. As shown in Fig.~\ref{fig:sI_evo_19}, the evolution of the fitness yielded only two high-fitness solutions. In unbiased self-assembly simulations, both of these solutions resulted in the formation of the sI clathrate rather than the intended SOD structure. Silica clathrates form a distinct subset of zeolites characterized by rings that are too small to allow the free movement of guest molecules within the crystal. However, it is precisely this property that makes them highly researched for gas storage applications. Molinero previously succeeded in reproducing sII clathrate (also known as MTN) in previous work~\cite{kumar_could_2018}.

\begin{figure}[H]
    \centering
    \includegraphics[width=1.0\linewidth]{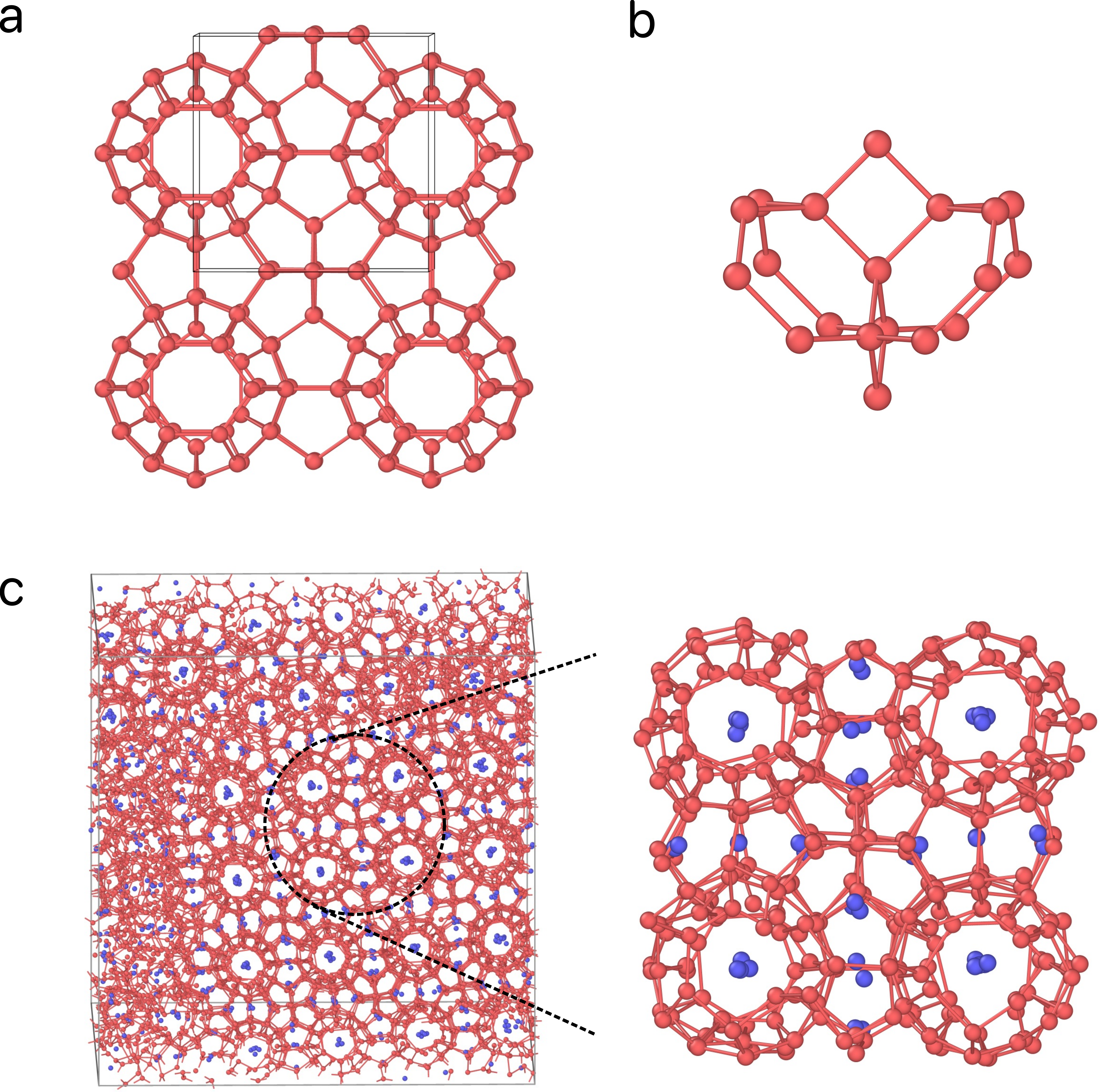}
    \caption{(a) The structure of the perfect sI clathrate. (b) A crystalline seed composed of 19 T particles. (c) The sI clathrate obtained from an unbiased self-assembly simulation using the high-fitness solution from the 22nd generation and a zoom-in diagram.}
    \label{fig:sI}
\end{figure}

\begin{table}[H]
    \setlength{\tabcolsep}{3pt}
    \centering
    \begin{tabular}{cccccccc}
    \hline
        Solution & $T$ & $\varepsilon_{\mathrm{TT}}$ & $\varepsilon_{\mathrm{TS}}$ & $\varepsilon_{\mathrm{SS}}$ & $\sigma_{\mathrm{TS}}$ & $\sigma_{\mathrm{SS}}$ & $\lambda$ \\
        in     & K   & eV    & eV    & eV    & \AA  & \AA  &       \\\hline
        Gen 22 & 645 & 0.404 & 0.079 & 0.059 & 4.37 & 7.98 & 16.99 \\
        Gen 23 & 650 & 0.394 & 0.078 & 0.051 & 4.46 & 8.68 & 18.25 \\
    \hline
    \end{tabular}
    \caption{The high-fitness solutions that spontaneously self-assemble into the sI clathrate within 50 ns.}
    \label{tab:sI_solutions}
\end{table}

The structure of the sI clathrate shows that there is one S particle in each cage, regardless of whether it is a small cage, $5^{12}$, or a large cage, $5^{12} 6^2$, as shown in Fig.~\ref{fig:sI}(c). Consequently, the T particle composition, $\chi_{\mathrm{T}}=46/(46+8)\approx0.85$, is similar to that of SOD. Compared to the parameters optimized for SOD, the main differences lie in $\varepsilon_{\mathrm{SS}}$ and $\lambda$. The lower value of $\varepsilon_{\mathrm{SS}}$ facilitates the formation of an anisotropic distribution of S particles in the clathrate structure.

%% file: results_cfi.tex
\subsection{Hole-type CFI zeolite}

\begin{figure}[H]
    \centering
    \includegraphics[width=1.0\linewidth]{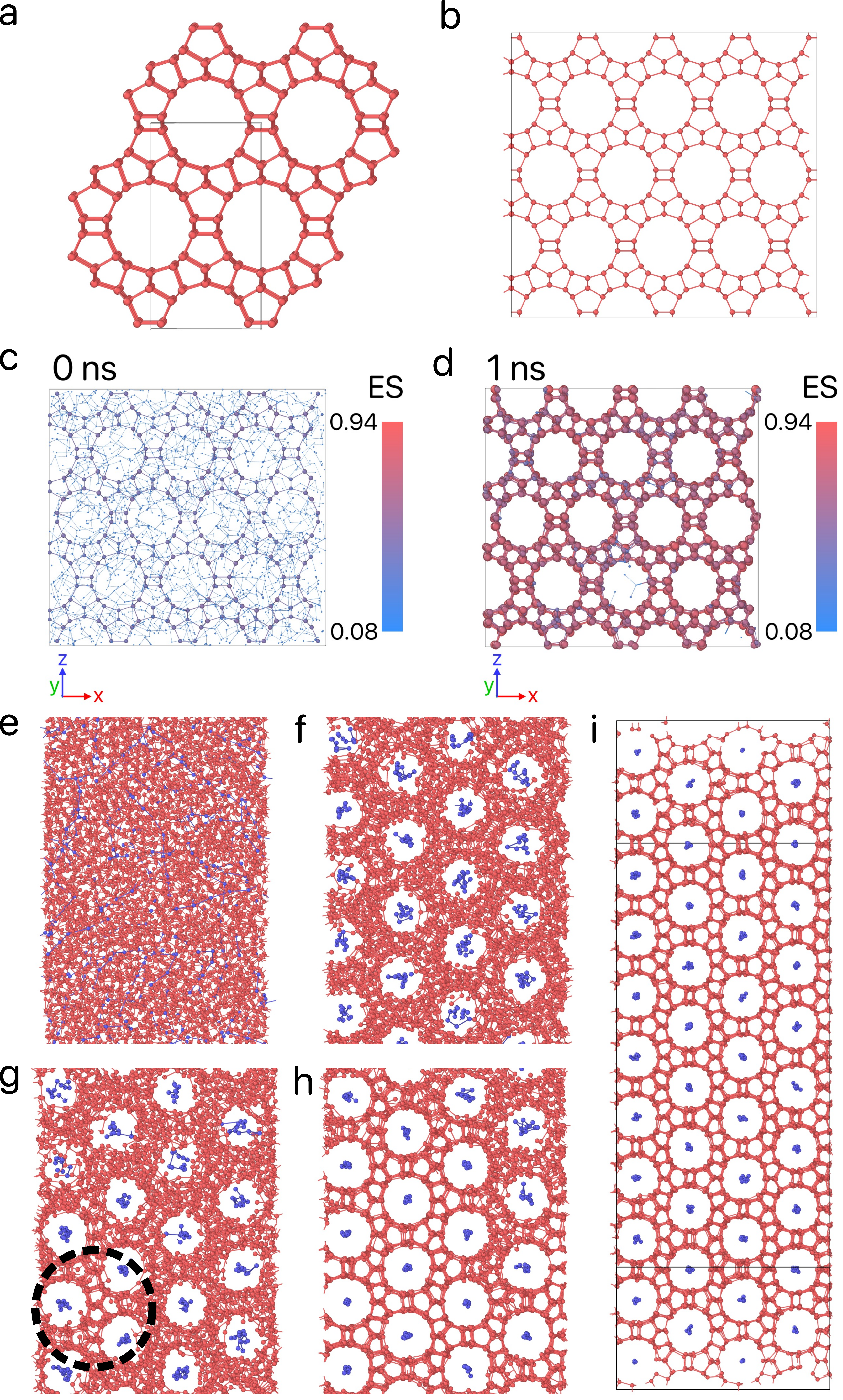}
    \caption{(a) A partial view of the perfect CFI structure. (b) A crystalline seed consisting of a single-particle-thick layer. (c) The initial configuration at 0 ns and (d) the final configuration at 1 ns of a seed-pinning simulation for the highest-fitness sample from the 30th generation of the inverse design protocol for CFI. The radius of the shown T particles is proportional to their environment similarity (ES) to the target framework, as indicated by the colorbar. (e-h) The CFI structure obtained from an unbiased self-assembly simulation using the highest-fitness sample from the 30th generation. 
    (e) An amorphous mixture of T and S particles. (f) A metastable hexagonal mesophase, where chains of S particles support the channels. (g) A nucleation site forms within this structure, highlighted by the dashed circle. (h) Growth of the CFI framework. (i) The fully developed CFI framework spans the entire system. To our knowledge, there are no previously reported parameters for CFI in the T-S model.}
    \label{fig:cfi}
\end{figure}

We now turn our attention to the inverse design of CFI within our coarse-grained T-S model. CFI is a porous framework with one-dimensional channels, i.e. a hole-type framework. To our knowledge, no T-S model parameter sets have been optimized for the self-assembly of CFI. The CFI unit cell consists of 32 T particles and includes 2 S particles, resulting in a composition of $\chi_{\mathrm{T}} = 32/(32+2) \approx 0.94$~\cite{cfi}. The structure can be characterized by slices perpendicular to the channel direction. Therefore, for the inverse design, we employ a crystalline seed consisting of a single-particle-thick layer that spans the entire simulation box rather than a spherical seed, as illustrated in Fig.~\ref{fig:cfi}(b). Furthermore, due to the significant differences in S-S distances within and between the channels, a lower value of $\varepsilon_{\mathrm{SS}}$, i.e. the strength of the isotropic two-body interaction between S particles, is required. Therefore, we reduce the optimization range of $\varepsilon_{\mathrm{SS}}$ from 0.02--0.09 to 0.01--0.05 eV. We again use the same inverse design recipe as described previously. 

\begin{figure}
    \centering
    \includegraphics[width=1.0\linewidth]{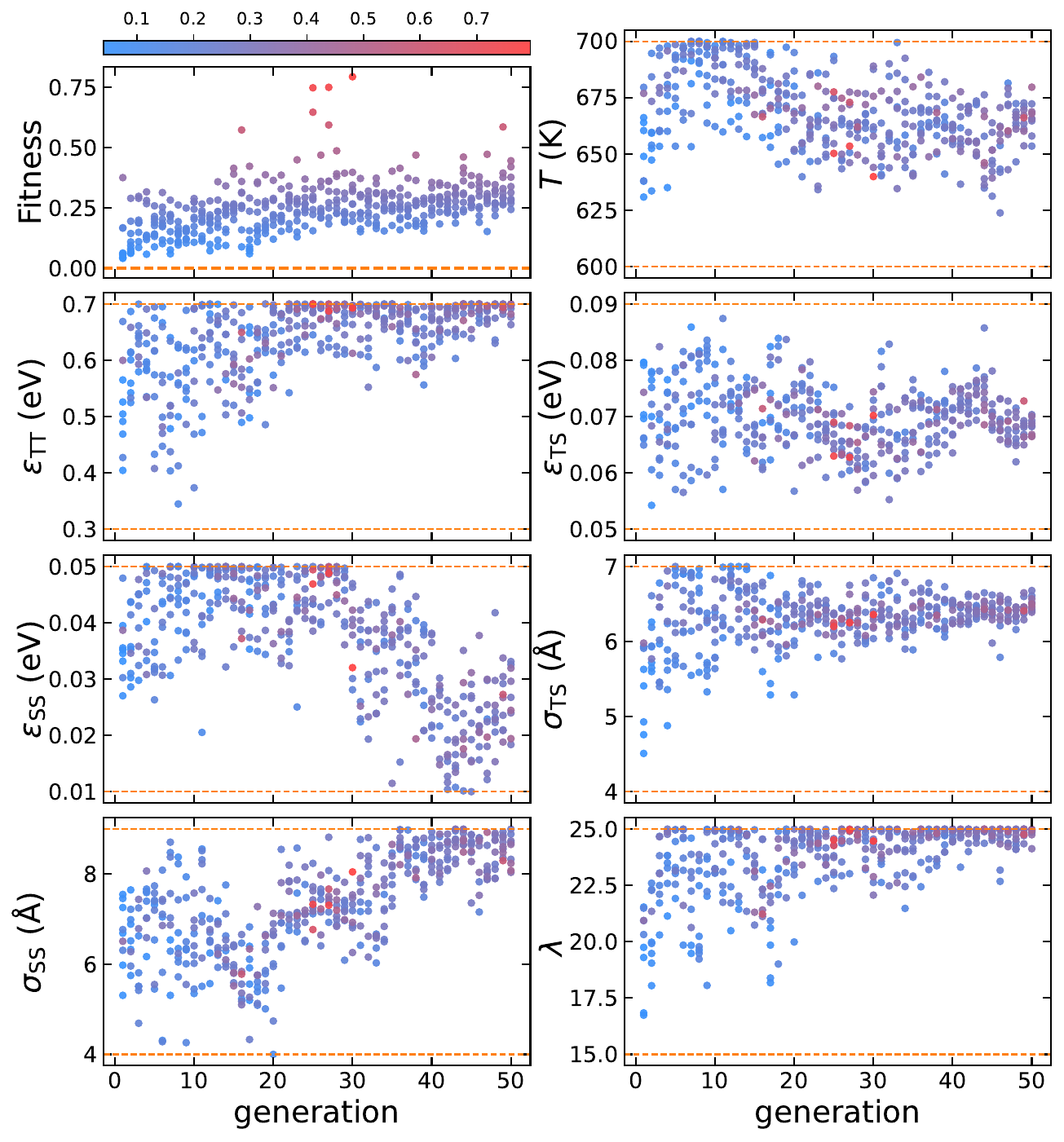}
    \caption{Evolution of the design parameters for CFI, with each point colored according to its fitness. The dashed orange lines represent the parameter boundaries.}
    \label{fig:cfi_evo_256}
\end{figure}

\begin{table}[H]
    \setlength{\tabcolsep}{3pt}
    \centering
    \begin{tabular}{cccccccc}
    \hline
        Solution & $T$ & $\varepsilon_{\mathrm{TT}}$ & $\varepsilon_{\mathrm{TS}}$ & $\varepsilon_{\mathrm{SS}}$ & $\sigma_{\mathrm{TS}}$ & $\sigma_{\mathrm{SS}}$ & $\lambda$ \\
        in     & K   & eV    & eV    & eV    & \AA  & \AA  &       \\\hline
        Gen 30 & 640 & 0.693 & 0.070 & 0.032 & 6.36 & 8.05 & 24.47 \\
    \hline
    \end{tabular}
    \caption{The high-fitness solution that spontaneously self-assembles into the hole-type CFI zeolite within 200 ns.}
    \label{tab:cfi_solutions}
\end{table}

We present the evolution of fitness values during the optimization process for CFI self-assembly in Fig.~\ref{fig:cfi_evo_256}. We observe a clear trend of increasing fitness with stronger tetrahedral interactions, as indicated by the continuous rise in both $\varepsilon_{\mathrm{TT}}$ and $\lambda$. This is consistent with the highly tetrahedral nature of the CFI framework. We select the highest fitness sample from the 30th generation, see Table~\ref{tab:cfi_solutions}. Compared to the interaction parameters for cage-type zeolites, e.g. SGT and SOD, a higher value of $\sigma_{\mathrm{TS}}\approx$ 6 \AA, which closely matches the radius of the largest 14-membered ring, provides strong geometric support for the CFI framework. An unbiased simulation using this parameter set successfully self-assembles into the CFI framework, revealing a two-step nucleation pathway for its crystallization, as shown in Figs.~\ref{fig:cfi}(e-i). As illustrated in Fig.~\ref{fig:cfi}(g), parallel strings of S particles initially guide the formation of a hexagonal mesophase, where T particles arrange into a locally disordered hexagonal structure that retains long-range order. A nucleation site subsequently forms within this structure. As shown in Fig.~\ref{fig:cfi}(i), the nucleus rapidly grows, transforming the entire crystal into the CFI framework within 1 ns (see Supplementary Materials). 

%% file: results_afi.tex
\subsection{Hole-type AFI zeolite and Polymorph Selection}

\begin{figure}[H]
    \centering
    \includegraphics[width=1.0\linewidth]{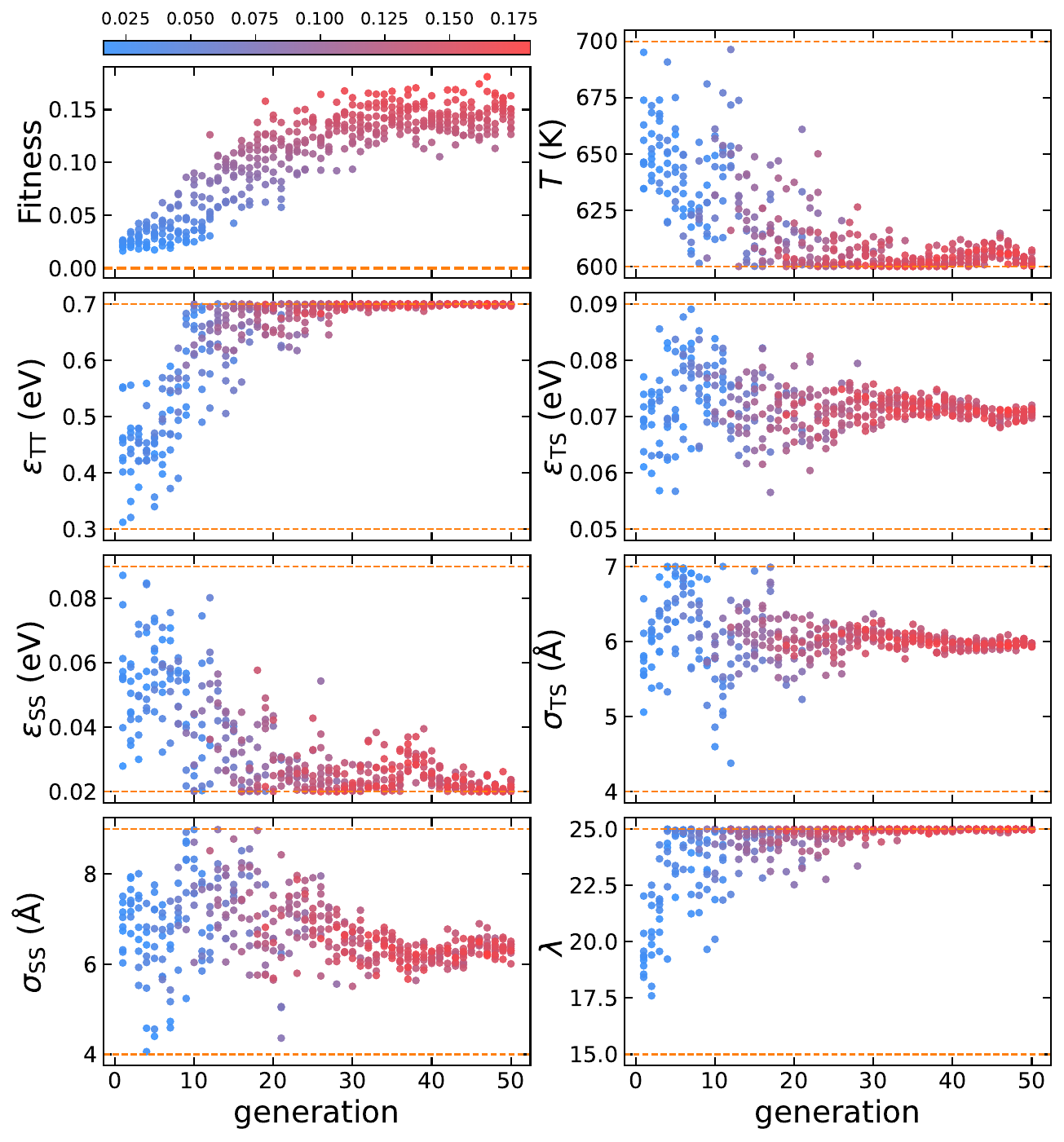}
    \caption{Evolution of the design parameters for the hole-type AFI zeolite framework, with each point colored according to its fitness. The dashed orange lines represent the parameter boundaries.}
    \label{fig:afi_evo_192}
\end{figure}

\begin{figure}[H]
    \centering
    \includegraphics[width=1.0\linewidth]{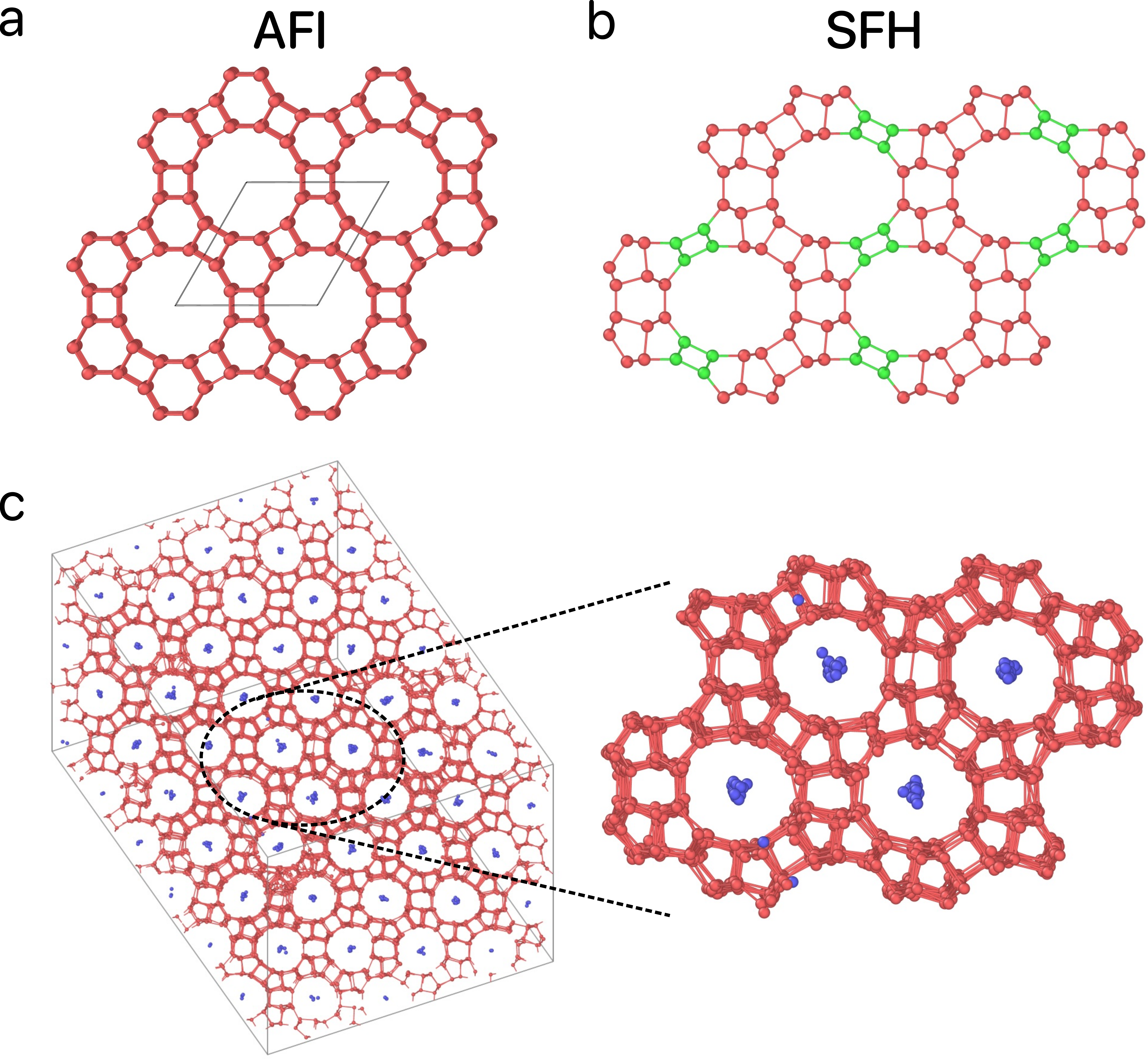}
    \caption{(a) The structure of the targeted AFI framework. (b) The structure of the SFH framework, which is the closest match to the new Z5 framework. The rings formed by green particles in the SFH framework are absent in the new Z5 framework. (c) The newly discovered Z5 framework, assembled using the highest-fitness solution from the 19th generation.}
    \label{fig:idk}
\end{figure}

As demonstrated by the discovery of the sI clathrate structure during the inverse design of the SOD zeolite, the unbiased self-assembly test can occasionally yield a polymorph distinct from the targeted framework when using seed-pinning. In this section, we report the discovery of an uncataloged framework, which we designate as Z5, during the inverse design of the hole-type AFI framework~\cite{afi}. AFI is a one-dimensional porous zeolite characterized by its large 12-membered ring as shown in Fig.~\ref{fig:idk}(a). The AFI unit cell contains 24 T particles and includes 2 S particles, resulting in a composition of $\chi_{\mathrm{T}} = 24/(24+2) \approx 0.92$. Similar to the CFI case, we use a crystalline seed of a single-particle-thick layer in our inverse design simulations. We follow again our inverse design protocol to optimize the seven design parameters for the self-assembly of AFI. The number of unique environments is $N_E=24$. We present the results in Fig.~\ref{fig:afi_evo_192}. The evolution of the design parameters, as shown in Fig.~\ref{fig:afi_evo_192}, suggests that some optimal parameter values may lie outside the predefined ranges, i.e. $T$, $\varepsilon_{\mathrm{TT}}$, $\varepsilon_{\mathrm{SS}}$, and $\lambda$ have reached upper or lower boundaries, and the fitness has reached a plateau. We selected the highest-fitness parameter combination, as presented in Table~\ref{tab:idk_solutions}.

When tested in an unbiased self-assembly simulation, this parameter set leads to the discovery of a new zeolite framework not cataloged in IZA, as illustrated in Fig.~\ref{fig:idk}(c). This new porous zeolite, which we label Z5, features the largest 12-membered ring, consistent with the target AFI framework.
However, the substructures within the T framework differ significantly. A more similar T substructure can be found in the IZA database, i.e. the SFH framework~\cite{sfh}, as shown in Fig.~\ref{fig:idk}(b). However, the key difference is that the SFH framework contains two additional rectangular rings along the large 12-membered ring, which are absent in our newly discovered Z5 framework. Furthermore, unlike the two-step nucleation pathway observed for the CFI framework, the crystallization of Z5 occurs without any intermediate mesophases.

The fact that the final structure obtained from unbiased self-assembly simulations differs from that obtained in biased seeding simulations, despite using the same parameters, suggests the existence of multiple nucleation pathways. These pathways may diverge or converge before or after reaching the seed size, which limits our control over growth in unbiased simulations. The diversity of space groups introduced by the tetrahedral many-body interactions also introduces potential competition between frameworks, particularly those with similar pore diameters, such as in AFI, CFI, and Z5.

\begin{table}
    \setlength{\tabcolsep}{3pt}
    \centering
    \begin{tabular}{cccccccc}
    \hline
        Solution & $T$ & $\varepsilon_{\mathrm{TT}}$ & $\varepsilon_{\mathrm{TS}}$ & $\varepsilon_{\mathrm{SS}}$ & $\sigma_{\mathrm{TS}}$ & $\sigma_{\mathrm{SS}}$ & $\lambda$ \\
        in     & K   & eV    & eV    & eV    & \AA  & \AA  &       \\\hline
        Gen 19 & 604 & 0.699 & 0.070 & 0.022 & 5.95 & 6.82 & 24.93 \\
    \hline
    \end{tabular}
    \caption{The high-fitness solution that spontaneously self-assembles into the new Z5 framework within 200 ns.}
    \label{tab:idk_solutions}
\end{table}

%% file: conclusion.tex
\section{CONCLUSION}

In conclusion, we developed and presented a novel inverse design workflow capable of identifying optimal interaction parameters and thermodynamic conditions for the self-assembly of complex zeolite frameworks within a coarse-grained model for silica and a structure-directing agent in combination with enhanced sampling molecular dynamics simulations. We optimize the design parameters iteratively using the covariance matrix adaptation evolution strategy, and use the seed-pinning technique implemented via steered molecular dynamics simulations to accelerate the sampling of nucleation events of the target zeolite. To evaluate the fitness of a sample, we monitor the growth of the target zeolite using the environment similarity order parameter with respect to the target framework. Using our inverse design workflow, we successfully optimized the design parameters for the spontaneous self-assembly of a known framework-type Z1 and a known cage-type SGT zeolite. Interestingly, the interaction parameters and thermodynamic conditions differ from those identified in previous work~\cite{kumar_two-step_2018, dhabal_coarse-grained_2021, wang_homochiral_2015, kumar_could_2018}. Additionally, we reproduced a new cage-type SOD and a new hole-type CFI zeolite as obtained from the IZA database. Remarkably, we also discovered the sI clathrate structure and a new zeolite framework not cataloged in the IZA database during the inverse design of the SOD and AFI zeolite, respectively. Thus, the unbiased self-assembly simulations can occasionally yield polymorphs distinct from the targeted frameworks in seed-pinning simulations.

There are also some limitations to this workflow. For instance, the choice of the shape and size of seeds for different frameworks is somewhat subjective, and the settings of $\kappa$ require fine-tuning after some trial runs to ensure that the seeds do not melt. In this study, we fixed the composition, which is reasonable for cage-like frameworks. However, for pore structures, the number of S particles and unit cells does not have a strict ratio. Especially considering that synthesizing zeolites also requires controlling the concentration of SDAs~\cite{gomez2018introduction}, including composition as a design parameter would be appropriate. From our self-assembly simulations, it became clear that higher adaptation schemes do not always guarantee better performance. Successful self-assembly can still occur with solutions that have lower fitness values. Additionally, the optimization process is stochastic, meaning that even with identical parameters and initial conditions, the algorithm may converge to different final solutions.

Our inverse design framework is inherently adaptable to other self-assembling systems such as other complex open crystals, metal-organic frameworks, quasicrystals, and liquid crystals. By adjusting custom design parameters, this protocol can be tailored to explore and achieve self-assembly in a variety of systems. These contributions underscore the robustness and versatility of our approach, paving the way for future research in self-assembling materials and enabling the development of new theoretical models and practical applications.

To summarize, our methodology not only enables the screening of synthesis protocols but also facilitates the discovery of hypothetical zeolites. More specifically, our inverse design protocol may give insights in the required coarse-grained interaction parameters of the structure directing agent, i.e. the size and attraction strength of the organic cation, to favor the self-assembly of a specific zeolite. Using a coarse-to-fine-grained mapping approach such as a generative adversarial network~\cite{stieffenhofer2020adversarial}, the precise molecular details of the structure directing agent may be obtained. This will be explored in future work.